\documentstyle[aps,epsfig]{revtex}
\newcommand{\comment}[1]{{}}
\newcommand{\key}[1]{{}}
\begin{document}

 \twocolumn[\hsize\textwidth\columnwidth\hsize\csname
 @twocolumnfalse\endcsname

\title{Loop Quantum Gravity\\
{\em (Review written for the electronic journal LIVING REVIEWS\/)}}

\author{
 Carlo Rovelli,\\
 {\it Department of Physics and Astronomy,} \\ 
 {\it University of Pittsburgh, Pittsburgh Pa 15260, USA}\\ 
 rovelli@pitt.edu
 }
\date{September 29, 1997}
\maketitle\vskip.5cm

                    {\bf Abstract}

The problem of finding the quantum theory of the gravitational field, 
and thus understanding what is quantum spacetime, is still open.  One 
of the most active of the current approaches is loop quantum gravity.  
Loop quantum gravity is a mathematically well-defined, 
non-perturbative and background independent quantization of general 
relativity, with its conventional matter couplings.  The research in 
loop quantum gravity forms today a vast area, ranging from 
mathematical foundations to physical applications.  Among the most 
significative results obtained are: (i) The computation of the 
physical spectra of geometrical quantities such as area and volume; 
which yields quantitative predictions on Planck-scale physics.  (ii) A 
derivation of the Bekenstein-Hawking black hole entropy formula.  
(iii) An intriguing physical picture of the microstructure of quantum 
physical space, characterized by a polymer-like Planck scale 
discreteness.  This discreteness emerges naturally from the quantum 
theory and provides a mathematically well-defined realization of 
Wheeler's intuition of a spacetime ``foam''.  Long standing open 
problems within the approach (lack of a scalar product, 
overcompleteness of the loop basis, implementation of reality 
conditions) have been fully solved.  The weak part of the approach is 
the treatment of the dynamics: at present there exist several 
proposals, which are intensely debated.  Here, I provide a general 
overview of ideas, techniques, results and open problems of this 
candidate theory of quantum gravity, and a guide to the relevant 
literature.

\vskip1cm
]

\section{Introduction}

The loop approach to quantum gravity is ten years old.  The first 
announcement of this approach was given at a conference in India in 
1987 \cite{RovelliSmolin87}.  This tenth anniversary is a good 
opportunity for attempting an assessment of what has and what has not 
been accomplished in these ten years of research and enthusiasm.

During these ten years, loop quantum gravity has grown into a wide 
research area and into a solid and rather well-defined tentative 
theory of the quantum gravitational field.  The approach provides a 
candidate theory of quantum gravity.  It provides a physical picture 
of Planck scale quantum geometry, calculation techniques, definite 
quantitative predictions, and a tool for discussing classical problems 
such as black hole thermodynamics.

We do not know whether this theory is physically correct or not.  
Direct or indirect experimental corroboration of the theory is 
lacking.  This is the case, unfortunately, for all present approaches 
to quantum gravity, due, of course, to the minuteness of the scale at 
which quantum properties of spacetime (presumably) manifest 
themselves.  In the absence of direct experimental guidance, we can 
evaluate a theory and compare it with alternative theories only in 
terms of internal consistency and consistency with what we do know 
about Nature.

Long standing open problems within the theory (such as the lack of a 
scalar product, the incompleteness of the loop basis and the related 
difficulty of dealing with identities between states, or the 
difficulty of implementing the reality conditions in the quantum 
theory) have been solidly and satisfactorily solved.  But while it 
is fairly well developed, loop quantum gravity is not yet a complete 
theory.  Nor has its consistency with classical general relativity 
been firmly established yet.  The sector of the theory which has not 
yet solidified is the dynamics, which exists in several variants 
presently under intense scrutiny.  On the other hand, in my opinion the 
strength of the theory is its compelling capacity to describe quantum 
spacetime in a background independent nonperturbative manner, and its 
genuine attempt to synthesize the conceptual novelties introduced by 
quantum mechanics with the ones introduced by general relativity.

The other large research program for a quantum theory of gravity, 
besides loop quantum gravity, is string theory, which is a tentative 
theory as well, and is more ambitious than loop gravity, since it also 
aims at unifying all known fundamental physics into a single theory.  
In section \ref{strings}, I will compare strengths and weaknesses of 
these two competing approaches to quantum gravity.

This ``living review'' is intended as a tool for orienting the 
reader in the field of loop gravity.  Here is the plan of the 
review:

\begin{itemize}

\item Section \ref{2}, \textbf{``Quantum Gravity: where are we?''}, is 
an introduction to the problem, the reason of its 
relevance, and the present state of our knowledge.

\item Section \ref{3}, \textbf{``History of loop quantum gravity''}, is 
a short overview of the historical development of the theory. 

\item Section \ref{3.5}, \textbf{``Resources''} contains pointers to 
introductory literature, institutions where loop gravity is 
studied, web pages, and other information that may be of use to 
students and researchers.

\item Section \ref{4}, \textbf{``Main ideas and physical inputs''}, 
discusses the physical and mathematical ideas on which loop 
quantum gravity is based, at a rather technical level.

\item The actual theory is described in detail in section \ref{5}, 
\textbf{``The formalism''}. 

\item Section \ref{6}, \textbf{``Results''}, is devoted to the results 
that have been derived from the theory.  I have divided results in two 
groups.  First, the ``technical'' results (\ref{technical}), namely the 
ones that have importance for the construction and the understanding 
of the theory itself, or that warrant the theory's consistency.  
Second, the ``physical'' results (\ref{physical}): what the theory says 
about the physical world.

\item In section \ref{7}, \textbf{``Open problems and current 
lines of investigation''}, I illustrate what I consider the main 
open problems, and the main currently active research lines.

\item In section \ref{8}, \textbf{``Short summary and 
conclusion''}, I summarize very briefly the state and the result 
of the theory, and present (necessarily very preliminary!)  
conclusions.

\end{itemize}

At the price of several repetitions, the structure of this review is very 
modular: sections are to a large extent independent from each other, 
have different style, and can be combined according to the interest of 
the reader.  A reader interested only in a very brief overview of the 
theory and its results, can find this in section \ref{8}.  Graduate 
students or persons of general culture may get a general idea of what 
goes on in this field and its main ideas from sections \ref{2} and 
\ref{6}.  If interested only in the technical aspects of the theory 
and its physical results, one can read sections \ref{5} and \ref{6} 
alone.  Scientists working in this field can use section \ref{5} and 
\ref{6} as a reference, and I hope they will find sections \ref{2}, 
\ref{3} and \ref{4} and \ref{7} stimulating\ldots

I will not enter technical details.  However, I will point to the 
literature where these details are discussed.  I have tried to be as 
complete as possible in indicating all relevant aspects and potential 
difficulties of the issues discussed.

The literature in this field is vast, and I am sure that there are 
works whose existence or relevance I have failed to recognize.  I 
sincerely apologize with the authors whose contributions I have 
neglected or under-emphasized and I strongly urge them to contact me 
to help me making this review more complete.  The ``living reviews'' 
are constantly updated and I should be able to correct errors and 
omissions.

\section{Quantum Gravity: where are we?}\label{2}

This is a non-technical section in which I illustrate the problem 
of quantum gravity in general, its origin, its importance, and 
the present state of our knowledge in this regard.

The problem of describing the quantum regime of the gravitational 
field is still open.  There are tentative theories, and competing 
research directions.  For an overview, see \cite{Isham95}.  The 
two largest research programs are string theory and loop quantum 
gravity.  But several other directions are being explored, such 
as twistor theory \cite{Twistors}, noncommutative geometry 
\cite{Noncommutative}, simplicial quantum gravity 
\cite{Simplicial,Simplicial2,Simplicial3,Simplicial4}, Euclidean 
quantum gravity \cite{Euclidean,Euclidean2}, the Null Surface 
Formulation \cite{Newman,Newman2,Newman3} and others.

String theory and loop quantum gravity differ not only because 
they explore distinct physical hypotheses, but also because they 
are expressions of two separate communities of scientists, which 
have sharply distinct prejudices, and view the problem of quantum 
gravity in surprisingly different manners.

\subsection{What is the problem?  The view of a high energy 
physicist}

High energy physics has obtained spectacular successes during this 
century, culminated with the (far from linear) establishment of 
quantum field theory as the general form of dynamics and with the 
comprehensive success of the $SU(3)\times SU(2)\times U(1)$ Standard 
Model.  Thanks to this success, now a few decades old, physics is in a 
condition in which it has been very rarely: there are no experimental 
results that clearly challenge, or clearly escape, the present 
fundamental theory of the world.  The theory we have encompasses 
virtually everything -- except gravitational phenomena.  From the 
point of view of a particle physicist, gravity is then simply the last 
and weakest of the interactions.  It is natural to try to understand 
its quantum properties using the strategy that has been so successful 
for the rest of microphysics, or variants of this strategy.  The 
search for a conventional quantum field theory capable of embracing 
gravity has spanned several decades and, through an adventurous 
sequence of twists, moments of excitement and disappointments, has 
lead to string theory.  The foundations of string theory are not yet 
well understood; and it is not yet entirely clear how a supersymmetric 
theory in 10 or 11 dimensions can be concretely used for deriving 
comprehensive univocal predictions about our world.\footnote{I heard 
the following criticism to loop quantum gravity: ``Loop quantum 
gravity is certainly physically wrong, because: (1) it is not 
supersymmetric, (2) is formulated in four dimensions''.  But 
experimentally, the world still insists on looking four-dimensional 
and not supersymmetric.  In my opinion, people should be careful of 
not being blinded by their own speculation, and mistaken interesting 
hypotheses (such as supersymmetry and high-dimensions) for established 
truth.} But string theory may claim extremely remarkable theoretical 
successes and is today the leading and most widely investigated 
candidate theory of quantum gravity.

In string theory, gravity is just one of the excitations of a string 
(or other extended object) living over some background metric space.  
The existence of such background metric space, over which the theory 
is defined, is needed for the formulation and for the interpretation 
of the theory, not only in perturbative string theory, but in the 
recent attempts of a non-perturbative definition of the theory, such 
as $M$ theory, as well, in my understanding.  Thus, for a physicist 
with a high energy background, the problem of quantum gravity is now 
reduced to an aspect of the problem of understanding what is the 
mysterious nonperturbative theory that has perturbative string theory 
as its perturbation expansion, and how to extract information on 
Planck scale physics from it.

\subsection{What is the problem? The view of a relativist}

For a relativist, on the other hand, the idea of a fundamental 
description of gravity in terms of physical excitations over a 
background metric space sounds physically very wrong.  The key lesson 
learned from general relativity is that there is no background metric 
\textit{over} which physics happens (unless, of course, in 
approximations).  The world is more complicated than that.  Indeed, 
for a relativist, general relativity is much more than the field 
theory of a particular force.  Rather, it is the discovery that 
certain classical notions about space and time are inadequate at the 
fundamental level; they require modifications which are possibly as 
basics as the ones that quantum mechanics introduced.  One of such 
inadequate notions is precisely the notion of a background metric 
space (flat or curved), over which physics happens.  This profound 
conceptual shift has led to the understanding of relativistic gravity, 
to the discovery of black holes, to relativistic astrophysics and to 
modern cosmology.

From Newton to the beginning of this century, physics has had a solid 
foundation in a small number of key notions such as space, time, 
causality and matter.  In spite of substantial evolution, these 
notions remained rather stable and self-consistent.  In the first 
quarter of this century, quantum theory and general relativity have 
modified this foundation in depth.  The two theories have obtained 
solid success and vast experimental corroboration, and can be now 
considered as established knowledge.  Each of the two theories 
modifies the conceptual foundation of classical physics in a (more or 
less) internally consistent manner, but we do not have a novel 
conceptual foundation capable of supporting \textit{both} theories.  
This is why we do not yet have a theory capable of predicting what 
happens in the physical regime in which both theories are relevant, 
the regime of Planck scale phenomena, $10^{-33}$ cm.

General relativity has taught us not only that space and time share 
the property of being dynamical with the rest of the physical 
entities, but also --more crucially-- that spacetime location is 
relational only (see section \ref{relational}).  Quantum mechanics has 
taught us that any dynamical entity is subject to Heisenberg's 
uncertainty at small scale.  Thus, we need {\em a relational notion of 
a quantum spacetime\/}, in order to understand Planck scale physics.

Thus, for a relativist, the problem of quantum gravity is the 
problem of bringing a vast conceptual revolution, started with 
quantum mechanics and with general relativity, to a conclusion 
and to a new synthesis.\footnote{For a detailed discussion of 
this idea, see \cite{RovelliHalf} and \cite{SmolinBook}.} In this 
synthesis, the notions of space and time need to be deeply 
reshaped, in order to keep into account what we have learned with 
both our present ``fundamental'' theories.

Unlike perturbative or nonperturbative string theory, loop 
quantum gravity is formulated without a background spacetime, and 
is thus a genuine attempt to grasp what is quantum spacetime at 
the fundamental level.  Accordingly, the notion of spacetime that 
emerges from the theory is profoundly different from the one on 
which conventional quantum field theory or string theory are 
based.

\subsection{Strings or loops?} \label{strings}

Above, I have emphasized the radically distinct cultural paths leading 
to string theory and loop quantum gravity.  Here, I attempt a 
comparison of the actual achievements that the two theories have 
obtained so far for what concerns the description of Planck scale 
physics.

Once more, however, I want to emphasize here that, whatever prejudices 
this or that physicist may have, both theories are {\em tentative\/}: 
as far as we really know, either, or both, theories could very well 
turn out to be physically entirely wrong.  And I do not mean that they 
could be superseded: I mean that all their specific predictions could 
be disproved by experiments.  Nature does not always share our 
aesthetic judgments, and the history of theoretical physics is full of 
enthusiasm for strange theories turned into disappointment.  The 
arbiter in science are experiments, and {\em not a single experimental 
result supports, not even very indirectly, any of the current theories 
that go beyond the Standard Model and general relativity}.  To the 
contrary, all the predictions made so far by theories that go beyond 
the Standard Model and general relativity (proton decay, 
supersymmetric particles, exotic particles, solar system dynamics) 
have for the moment been punctually falsified by the experiments.  
Comparing this situation with the astonishing experimental success of 
the Standard Model and classical general relativity should make us 
very cautious, I believe.  Lacking experiments, theories can only be 
compared on completeness and aesthetic criteria -- criteria, one 
should not forget, that according to many favored Ptolemy over 
Copernicus, at some point.

The main merit of string theory is that it provides a superbly elegant 
unification of known fundamental physics, and that it has a well 
defined perturbation expansion, finite order by order.  Its main 
incompleteness is that its non-perturbative regime is poorly 
understood, and that we do not have a background-independent 
formulation of the theory.  In a sense, we do not really know what is 
the theory we are talking about.  Because of this poor understanding 
of the non perturbative regime of the theory, Planck scale physics and 
genuine quantum gravitational phenomena are not easily controlled: 
except for a few computations, there is not much Planck scale physics 
derived from string theory so far.   There are, however, two sets of 
remarkable physical results.  The first is given by some very high 
energy scattering amplitudes that have been computed (see for instance 
\cite{AmatiVeneziano,AmatiVeneziano2,%
AmatiVeneziano3,AmatiVeneziano4,Verlinde,t'Hooft}).  An intriguing 
aspect of these results is that they indirectly suggest that geometry 
below the Planck scale cannot be probed --and thus in a sense does not 
exist-- in string theory.  The second physical achievement of string 
theory (which followed the d-branes revolution) is the recent 
derivation of the Bekenstein-Hawking black hole entropy formula for 
certain kinds of black holes 
\cite{StromingerVafa,HorowitzStrominger,HorowitzEtAl1,HorowitzEtAl2}.

The main merit of loop quantum gravity, on the other hand, is 
that it provides a well-defined and mathematically rigorous 
formulation of a background-independent non-perturbative 
generally covariant quantum field theory.  The theory provides a 
physical picture and quantitative predictions on the world 
at the Planck scale.  The main incompleteness of the theory 
regards the dynamics, formulated in several variants.  So far, 
the theory has lead to two main sets of physical results.  The 
first is the derivation of the (Planck scale) eigenvalues of 
geometrical quantities such as areas and volumes. The second is
the derivation of black hole entropy for ``normal'' black holes 
(but only up to the precise numerical factor).  

Finally, strings and loop gravity, may not necessarily be competing 
theories: there might be a sort of complementarity, at least 
methodological, between the two.  This is due to the fact that the 
open problems of string theory regard its background-independent 
formulation, and loop quantum gravity is precisely a set of techniques 
for dealing non-perturbatively with background independent theories.  
Perhaps the two approaches might even, to some extent, converge.  
Undoubtedly, there are similarities between the two theories: first of 
all the obvious fact that both theories start with the idea that the 
relevant excitations at the Planck scale are one dimensional objects 
-- call them loops or strings.  I understand that in another living 
review in this issue, Lee Smolin explores the possible relations 
between string theory and loop gravity \cite{Smolin97}.

\section{History of loop quantum gravity, main steps}\label{3}

The following chronology does not exhaust the literature on loop 
quantum gravity.  It only indicates the key steps in the construction 
of the theory, and the first derivation of the main results.  For more 
complete references, see the following sections.  (In the effort 
of grouping some results, something may appear a bit out of the 
chronological order.) 

\begin{description}

	\item[1986] {\bf  Connection formulation of classical general 
 	relativity\\ {\em Sen, Ashtekar.}}\\ 
	Loop quantum gravity is based on the formulation of classical 
	general relativity, which goes under the name of ``new 
	variables'', or ``Ashtekar variables'', or ``connectio-dynamics'' 
	(in contrast to Wheeler's ``geometro-dynamics'').  In this 
	formulation, the field variable is a self-dual connection, instead 
	of the metric, and the canonical constraints are simpler than in 
	the old metric formulation.  The idea of using a self-dual 
	connection as field variable and the simple constraints it yields 
	were discovered by Amitaba Sen \cite{Sen}.  Abhay Ashtekar 
	realized that in the $SU(2)$ extended phase space a self-dual 
	connection and a densitized triad field form a canonical pair 
	\cite{Ashtekar86,Ashtekar87} and set up the canonical formalism 
	based on such pair, which is the Ashtekar formalism.  Recent works 
	on the loop representation are not based on the original 
	Sen-Ashtekar connection, but on a real variant of it, whose use 
	has been introduced into Lorentzian general relativity by 
	\textbf{\em Barbero} \cite{Barbero,Barbero3,Barbero2,Barbero4}.

	\item[1986] {\bf  Wilson loop solutions of the hamiltonian 
	constraint\\ {\em Jacobson, Smolin}}.\\ Soon after 
	the introduction of the classical Ashtekar variables, Ted 
	Jacobson and Lee Smolin realized in \cite{JacobsonSmolin} 
	that the Wheeler-DeWitt equation reformulated in terms of the 
	new variables admits a simple class of exact solutions: the 
	traces of the holonomies of the Ashtekar connection around 
	smooth non-selfintersecting loops.  In other words: the 
	Wilson loops of the Ashtekar connection solve the 
	Wheeler-DeWitt equation if the loops are smooth and non 
	self-intersecting.

	\item[1987] {\bf  The Loop Representation\\ 
	{\em Rovelli, Smolin}}. \\
	The discovery of the Jacobson-Smolin Wilson loop solutions 
	prompted Carlo Rovelli and Lee Smolin 
	\cite{RovelliSmolin87,Rovelli88,RovelliSmolin88,RovelliSmolin90} 
	to ``change basis in the Hilbert space of the theory'', choosing 
	the Wilson loops as the new basis states for quantum gravity.  
	Quantum states can be represented in terms of their expansion on 
	the loop basis, namely as functions on a space of loops.  This 
	idea is well known in the context of canonical lattice Yang-Mills 
	theory \cite{WilsonKoguth}, and its application to continuous 
	Yang-Mills theory had been explored by \textbf{\em Gambini and 
	Trias} \cite{GambiniTrias2,GambiniTrias}, who developed a continuous 
	``loop representation'' much before the Rovelli-Smolin one.  The 
	difficulties of the loop representation in the context of 
	Yang-Mills theory are cured by the diffeomorphism invariance of GR 
	(see section \ref{diffinvariance} for details).  The loop 
	representation was introduced by Rovelli and Smolin as a 
	representation of a classical Poisson algebra of ``loop 
	observables''.  The relation with the connection representation 
	was originally derived in the form of an integral transform (an 
	infinite dimensional analog of a Fourier transform) from 
	functionals of the connection to loop functionals.  Several years 
	later, this loop transform was shown to be mathematically 
	rigorously defined \cite{AshtekarIsham}.  The immediate results of 
	the loop representation were two: the diffeomorphism constraint is 
	completely solved by knot states (loop functionals that depend 
	only on the knotting of the loops), making earlier suggestions by 
	Smolin on the role of knot theory in quantum gravity 
	\cite{Smolin88} concrete; and (suitable 
	\cite{RovelliSmolin90,Smolin93} extensions of) the knot states 
	with support on non-selfintersecting loops were proven to be 
	solutions of all quantum constraints, namely exact physical states 
	of quantum gravity.
	
	\item[1988 - ] {\bf Exact states of quantum gravity\\
	{\em Husain, Br\"ugmann, Pullin, Gambini, Kodama}}.\\
	The investigation of exact solutions of the quantum constraint 
	equations, and their relation with knot theory (in particular with 
	the Jones polynomial and other knot invariants) has started soon 
	after the formulation of the theory and continued since 
	\cite{H1,H5,H4,H6,H3,H7,H8,GambiniPullin,Kodama,PullinEtAl97}.

	\item[1989 - ] {\bf  Model theories\\
	{\em Ashtekar, Husain, Loll, Marolf, Rovelli, Samuel, Smolin, 
	Lewandowski, Marolf, Thiemann}}.\\
	The years immediately following the discovery of the loop 
	formalism were mostly dedicated to understanding 
	the loop representation by studying it in simpler contexts, 
	such as 2+1 general relativity 
	\cite{AshtekarEtAl89,Marolf93,AshtekarLoll}, Maxwell 
	\cite{AshtekarRovelli}, linearized gravity 
	\cite{AshtekarRovelliSmolin}, and, much later,  
	2d Yang-Mills theory \cite{ALMMT}.

	\item[1992] {\bf  Classical limit: weaves\\
	{\em Ashtekar, Rovelli, Smolin}}.\\
	The first indication that the theory predicts Planck scale 
	discreteness came from studying the states that approximate 
	geometries flat on large scale \cite{weave}.  These states, 
	denoted ``weaves'', have a ``polymer'' like structure at short 
	scale, and can be viewed as a formalization of Wheeler's 
	``spacetime foam''.

    \item[1992] {\bf  $C^{*}$ algebraic framework\\
	{\em Ashtekar, Isham}}.\\
	In \cite{AshtekarIsham}, Abhay Ashtekar and Chris Isham 
	showed that the loop transform introduced in gravity by 
	Rovelli and Smolin could be given a rigorous mathematical 
	foundation, and set the basis for a mathematical 
	systematization of the loop ideas, based on $C^{*}$ algebra 
	ideas.
		
	\item[1993] {\bf  Gravitons as embroideries over the weave\\
	{\em Iwasaki, Rovelli}}.\\
    In \cite{IwasakiRovelli} Junichi Iwasaki and Rovelli studied 
    the representation of gravitons in loop quantum gravity. 
    These appear as topological modifications of the fabric of 
    the spacetime weave. 

	\item[1993 - ] {\bf Alternative versions\\
	{\em Di Bartolo, Gambini, Griego, Pullin}}.\\
    Some versions of the loop quantum gravity alternative to the 
    ``orthodox'' version have been developed. In particular, these  
    authors above have developed the so called ``extended'' loop 
    representation. See \cite{Extended,Extended2}. 

	\item[1994 -] {\bf  Fermions,\\
	{\em Morales-Tecotl, Rovelli}}.\\  
	Matter coupling were beginning to be explored in 
	\cite{MoralesRovelli,MoralesRovelli2}.  Later, matter's kinematics 
	was studied by \textbf{\em Baez and Krasnov} 
	\cite{KrasnovMatter,BaezKrasnov}, while \textbf{\em Thiemann\/} 
	has extended his results on the dynamics to the coupled Einstein 
	Yang-Mills system in \cite{ThiemannYM}.

	\item[1994] {\bf  The $d\mu_{0}$ measure and the scalar product\\
	{\em Ashtekar, Lewandowski, Baez}}.\\  
	In \cite{AshtekarLewandowski,AshtekarLewandowski3,AshtekarLewandowski2} 
	Ashtekar and Lewandowski set the basis of the differential 
	formulation of loop quantum gravity by constructing its two key 
	ingredients: a diffeomorphism invariant measure on the space of 
	(generalized) connections, and the projective family of Hilbert 
	spaces associated to graphs.  Using these techniques, they were 
	able to give a mathematically rigorous construction of the state 
	space of the theory, solving long standing problems deriving from 
	the lack of a basis (the insufficient control on the algebraic 
	identities between loop states).  Using this, they defined a 
	consistent scalar product and proved that the quantum operators in 
	the theory were consistent with all identities.  John Baez showed 
	how the measure can be used in the context of conventional 
	connections, extended it to the non-gauge invariant states 
	(allowing the $E$ operator to be defined) and developed the use of 
	the graph techniques \cite{Baez94a,Baez93,Baez2}.  Important 
	contributions to the understanding of the measure were also given 
	by \textbf{\em Marolf and Mour\~ao} \cite{MarolfMourao}.
	
	\item[1994] {\bf Discreteness of area and volume eigenvalues\\
	{\em Rovelli, Smolin}}.\\
	In my opinion, the most significative result of loop quantum 
	gravity is the discovery that certain geometrical quantities, 
	in particular area and volume, are represented by operators 
	that have discrete eigenvalues.  This was found by Rovelli 
	and Smolin in \cite{RovelliSmolin95}, where the first set of 
	these eigenvalues were computed.   Shortly after, this result 
	was confirmed and extended by a number of authors, using very 
	diverse techniques.  In particular, \textbf{\em Renate Loll} 
	\cite{Loll95b,Loll95bb} used lattice techniques to analyze 
	the volume operator and corrected a numerical error in 
	\cite{RovelliSmolin95}.  \textbf{\em Ashtekar and 
	Lewandowski} \cite{LewandowskiArea,AshtekarLewandowskiArea2} 
	recovered and completed the computation of the spectrum of 
	the area using the connection representation, and new 
	regularization techniques.  \textbf{\em 
	Frittelli, Lehner and Rovelli} \cite{FrittelliEtAl} recovered 
	the Ashtekar-Lewandowski terms of the spectrum of the area, 
	using the loop representation.  \textbf{\em DePietri and 
	Rovelli\/} \cite{DePietriRovelli} computed general 
	eigenvalues of the volume.  Complete understanding of the 
	precise relation between different versions of the volume 
	operator came from the work of Lewandowski 
	\cite{Lewandowski97}.

	\item[1995] {\bf  Spin networks - solution of the overcompleteness 
	problem\\
	{\em Rovelli, Smolin, Baez}}.\\
	A long standing problem with the loop basis was its 
	overcompleteness.  A technical, but crucial step in understanding 
	the theory has been the discovery of the spin-network basis, which 
	solves this overcompleteness.  This step was taken by Rovelli and 
	Smolin in \cite{RovelliSmolin95b} and was motivated by the work of 
	\textbf{\em Roger Penrose} \cite{Penrose,Penrose2}, by analogous 
	bases used in lattice gauge theory and by ideas of Lewandowski 
	\cite{JerzyGraph}.  Shortly after, the spin network formalism was 
	cleaned up and clarified by \textbf{\em John Baez} 
	\cite{Baez95a,Baez95aa}.  After the introduction of the spin 
	network basis, all problems deriving from the incompleteness of 
	the loop basis are trivially solved, and the scalar product could 
	be defined also algebraically \cite{DePietriRovelli}.
	
	\item[1995] {\bf  Lattice\\
	{\em Loll, Reisenberger, Gambini, Pullin}}.\\
	Various lattice versions of the theory have appeared in 
	\cite{Loll95a,Reisenberger,GambiniPullin,GambiniLattice}.

	\item[1995] {\bf Algebraic formalism / Differential 
	formalism\\
	{\em DePietri Rovelli / Ashtekar, Lewandowski, Marolf, 
	Mour\~{a}o, Thiemann}}.\\
	The cleaning and definitive setting of the two main versions of 
	the formalisms was completed in \cite{DePietriRovelli} for the 
	algebraic formalism (the direct descendent of the old loop 
	representation); and in \cite{AshtekarEtAl95} for the differential 
	formalism (based on the Ashtekar-Isham $C^{*}$ algebraic 
	construction, on the Ashtekar-Lewandowski measure, on Don Marolf's 
	work on the use of formal group integration for solving the 
	constraints \cite{MarolfGroup,MarolfGroup2,MarolfGroup3}, and on 
	several mathematical ideas by Jos\'{e} Mour\~ao).
	
	\item[1996] {\bf Equivalence of the algebraic and 
	differential formalisms\\
	{\em DePietri}}.\\ 
	In \cite{DePietri}, Roberto DePietri proved the equivalence 
	of the two formalisms, using ideas from Thiemann 
	\cite{Thiemann95} and Lewandowski \cite{Lewandowski97}.

	\item[1996] {\bf Hamiltonian constraint\\
	{\em Thiemann}}.\\
	The first version of the loop hamiltonian constraint is in 
	\cite{RovelliSmolin88,RovelliSmolin90}.  The definition of 
	the constraint has then been studied and modified repeatedly, 
	in a long sequence of works, by \textbf{\em Br\"ugmann, 
	Pullin, Blencowe, Borissov} and others 
	\cite{H1,H2,H3,H4,H5,H6,H7,H8,H9}.  An important step was 
	made by Rovelli and Smolin in \cite{RovelliSmolin94} with the 
	realization that certain regularized loop operators have 
	finite limits on knot states (see \cite{LewandowskiMarolf}).  
	The search culminated with the work of Thomas Thiemann, who 
	was able to construct a rather well-defined hamiltonian 
	operator whose constraint algebra closes 
	\cite{Thiemann96,Thiemann96b,Thiemann96c}.  Variants of this 
	constraint have been suggested in 
	\cite{Smolin96,ReisenbergerRovelli} and elsewhere.

    \item[1996] {\bf Real theory: solution of the reality conditions 
    problem\\
    {\em Barbero, Thiemann}}.\\
	As often stressed by Karel Kucha\v{r}, implementing the 
	complicated reality condition of the complex connection into the 
	quantum theory was, until 1996, the main open problem in the loop 
	approach.\footnote{``The loop people have a credit card called 
	{\em reality conditions\/}, and whenever they solve a problem, 
	they charge the card, but one day the bill comes and the whole 
	thing breaks down like a card house'' \cite{kuchar}.} Following 
	the directions advocated by Fernando Barbero 
	\cite{Barbero,Barbero3,Barbero2,Barbero4}, namely to use the {\em 
	real\/} connection in the {\em Lorentzian\/} theory, Thiemann 
	found an elegant elegant way to completely bypass the problem.

	\item[1996] {\bf Black hole entropy\\
	{\em Krasnov, Rovelli}}.\\
	A derivation of the Bekenstein-Hawking formula for the 
	entropy of a black hole from loop quantum gravity was 
	obtained in \cite{Rovelli96}, on the basis of the ideas of 
	Kirill Krasnov \cite{Krasnov,Krasnov2}.  Recently, 
	\textbf{\em Ashtekar, Baez, Corichi and Krasnov} have 
	announced an alternative derivation \cite{AshtekarEtAl97}.

	\item[1997] {\bf  Anomalies\\
	{\em Lewandowski, Marolf, Pullin, Gambini}}.\\
	These authors have recently completed an extensive analysis of the 
	issue of the closure of the quantum constraint algebra and its 
	departures from the corresponding classical Poisson algebra 
	\cite{LewandowskiMarolf,LewandowskiEtAl}, following earlier 
	pioneering work in this direction by \textbf{\em Br\"ugmann, 
	Pullin, Borissov} and others \cite{BruegmannAlgebra,%
	BruegmannPullinAlgebra,GambiniGaratPullin,Borissov97}.  This 
	analysis has raised worries that the classical limit of 
	Thiemann's hamiltonian operator might fail to yield classical 
	general relativity, but the matter is still controversial.

	\item[1997] {\bf Sum over surfaces\\
	{\em Reisenberger Rovelli}}.\\  
	A ``sum over histories'' spacetime formulation of loop 
	quantum gravity was derived in 
	\cite{RovelliSurf,ReisenbergerRovelli} from the canonical 
	theory.  The resulting covariant theory turns out to be a sum 
	over topologically inequivalent surfaces, realizing earlier 
	suggestions by \textbf{\em Baez\/} 
	\cite{Baez,Baez2,Baez95a,Baez4}, Reisenberger  
	\cite{Reisenberger,Reisenberger94} and \textbf{\em Iwasaki\/} 
	\cite{Iwasaki} that a covariant version of loop gravity 
	should look like a theory of surfaces.  \textbf{\em Baez} has 
	studied the general structure of theories defined in this 
	manner \cite{Baez97}.  \textbf{\em Smolin and Markoupolou} 
	have explored the extension of the construction to the 
	Lorentzian case, and the possibility of altering the spin 
	network evolution rules \cite{SmolinMarkopoulo}.

\end{description}

\section{Resources}\label{3.5}

\begin{itemize}

\item A valuable resource for finding relevant literature is the 
comprehensive {\em ``Bibliography of Publications related to Classical 
and Quantum Gravity in terms of Connection and Loop Variables''}, 
organized chronologically.  The original version was compiled by 
Peter H\"ubner in 1989.  It has subsequently been updates by 
Gabriella Gonzales, Bernd Br\"ugmann, Monica Pierri and Troy 
Shiling.  Presently, it is being kept updated by Christopher 
Beetle and Alejandro Corichi.  The last version can be found on 
the net in \cite{Bibliography}.

\item This living review may serve as an updated introduction to 
quantum gravity in the loop formalism. More detailed (but less 
up to date) presentations are listed below. 
\begin{itemize}

\item {\em Ashtekar's book} \cite{AshtekarBook} may serve as a valuable 
basic introductory course on Ashtekar variables, particularly for 
relativists and mathematicians.  The part of the book on the loop 
representation is essentially an authorized reprint of parts of 
{\em the original Rovelli Smolin article} \cite{RovelliSmolin90}.  For 
this quantum part, I recommend looking at the article, rather 
than the book, because the article is more complete.

\item A simpler and more straightforward introduction to the 
Ashtekar variables and basic loop ideas can be found in the 
{\em Rovelli's review paper} \cite{Rovelli91}.  This is more oriented 
to a reader with physics background.

\item A recent general introduction to the new variables which 
includes several of the recent mathematical developments in the 
quantum theory is given by {\em Ashtekar's Les Houches 1992 lectures}  
\cite{AshtekarLesHouches}.

\item A particularly interesting collection of papers can be 
found in {\em the volume \cite{BaezBook1} edited by John Baez}.  The 
other {\em book by Baez, and Muniain} \cite{BaezBook2}, is a simple and 
pleasant introduction to several ideas and techniques in the 
field.

\item The last and up to date book on the loop representation is 
{\em the book by Gambini and Pullin} \cite{GambiniPullinBook}, 
especially good in lattice techniques and in the variant of loop 
quantum gravity called the ``extended loop representation'' 
\cite{Extended,Extended2} (which is nowadays a bit out of fashion, 
but remains an intriguing alternative to ``orthodox'' loop quantum 
gravity).

\item The two standard references for a complete presentation of 
the basics of the theory are: {\em DePietri and Rovelli 
\cite{DePietriRovelli} for the algebraic formulation}; and 
{\em Ashtekar, Lewandowski, Marolf, Mourao and Thiemann ($ALM^2T$) 
\cite{AshtekarEtAl95} for the differential formulation}. 

\end{itemize}

\item Besides the many conferences on gravity, the loop gravity 
community has met twice in Warsaw, in the {\em ``Workshop on Canonical 
and Quantum Gravity''}.  Hopefully, this will become a recurrent  
meeting.  This may be the right place to go for learning what is 
going on in the field.  For an informal account of the last of 
these meetings (August 1997), see \cite{RovelliMoG}.

\item Some of the main institutions where loop quantum gravity is 
studied are

\begin{itemize}

\item The {\em ``Center for Gravity and Geometry''} at Penn State 
University, USA. I recommend their invaluable {\em web page 
\cite{PennState}, maintained by Jorge Pullin}, for finding 
anything you need from the web.

\item  \emph{Pittsburgh University}, USA \cite{Pittsburgh}.

\item \emph{University of California at Riverside}, USA. John Baez 
moderates an interesting newsgroup, \emph{sci.physics.research}, 
with news from the field.  See \cite{John}.

\item \emph{Albert Einstein Institute}, Potsdam, Berlin, Europe 
\cite{Potsdam}.

\item \emph{Warsaw University}, Warsaw, Europe.

\item \emph{Imperial College}, London, Europe \cite{Imperial}.

\item \emph{Syracuse University}, USA \cite{Syracuse}.

\item \emph{Montevideo University}, Uruguay. 

\end{itemize}

\end{itemize}

\section{Main ideas and physical inputs}\label{4}

The main physical hypotheses on which loop quantum gravity relies 
are only general relativity and quantum mechanics.  In other 
words, loop quantum gravity is a rather conservative 
``quantization'' of general relativity,  with its 
traditional matter couplings.  In this sense, it is very 
different from string theory, which is based on a strong physical 
hypothesis with no direct experimental support (the world 
is made by strings).  

Of course ``quantization'' is far from a univocal algorithm, 
particularly for a nonlinear field theory.  Rather, it is a poorly 
understood inverse problem (find a quantum theory with the given 
classical limit).  More or less subtle choices are made in 
constructing the quantum theory.  I discuss these choices below.

\subsection{Quantum field theory on a differentiable manifold}

The main idea beyond loop quantum gravity is to take general 
relativity seriously.  We have learned with general relativity 
that the spacetime metric and the gravitational field are the same 
physical entity.  Thus, a quantum theory of the gravitational 
field is a quantum theory of the spacetime metric as well.  It 
follows that quantum gravity cannot be formulated as a quantum 
field theory over a {\em metric\/} manifold, because there is no 
(classical) metric manifold whatsoever in a regime in which 
gravity (and therefore the metric) is a quantum variable.

One could conventionally split the spacetime metric into two 
terms: one to be consider as a background, which gives a metric 
structure to spacetime; the other to be treated as a fluctuating 
quantum field.  This, indeed, is the procedure on which old 
perturbative quantum gravity, perturbative strings, as well as 
current non-perturbative string theories (M-theory), are based.  
In following this path, one assumes, for instance, that the 
causal structure of spacetime is determined by the underlying 
background metric alone, and not by the full metric.  Contrary to 
this, in loop quantum gravity we assume that the identification 
between the gravitational field and the metric-causal structure 
of spacetime holds, and must be taken into account, in the 
quantum regime as well.  Thus, no split of the metric is made, 
and there is no background metric on spacetime.

We can still describe spacetime as a (differentiable) manifold (a 
space without metric structure), over which quantum fields are 
defined.  A classical metric structure will then be defined by 
expectation values of the gravitational field operator.  Thus, 
the problem of quantum gravity is the problem of understanding 
what is {\em a quantum field theory on a manifold}, 
as opposed to quantum field theory on a metric space.  This is 
what gives quantum gravity its distinctive flavor, so different 
than ordinary quantum field theory.  In all versions of ordinary 
quantum field theory, the metric of spacetime plays an essential 
role in the construction of the basic theoretical tools (creation 
and annihilation operators, canonical commutation relations, 
gaussian measures, propagators \ldots ); these tools cannot be 
used in quantum field over a manifold.

Technically, the difficulty due to the absence of a background 
metric is circumvented in loop quantum gravity by defining the 
quantum theory as a representation of a Poisson algebra of 
classical observables which can be defined without using a 
background metric.  The idea that the quantum algebra at the basis 
of quantum gravity is not the canonical commutation relation 
algebra, but the Poisson algebra of a different set of observables 
has long been advocated by Chris Isham \cite{IshamLesHouches}, 
whose ideas have been very influential in the birth of loop 
quantum gravity.\footnote{Loop Quantum Gravity is an attempt to 
solve the last problem in Isham's lectures 
\cite{IshamLesHouches}.} The algebra on which loop gravity is the 
loop algebra \cite{RovelliSmolin90}.  The particular choice of 
{\em this\/} algebra is not harmless, as I discuss below.

\subsection{One additional assumption} \label{Additional}

In choosing the loop algebra as the basis for the quantization, 
we are essentially assuming that Wilson loop operators are well 
defined in the Hilbert space of the theory.  In other words, that 
certain states concentrated on one dimensional structures (loops 
and graphs) have finite norm.  This is a subtle non trivial 
assumptions entering the theory.  It is the key assumption that 
characterizes loop gravity.  If the approach turned out to be 
wrong, it will likely be because this assumption is wrong.  The 
Hilbert space resulting from adopting this assumption is not a 
Fock space.  Physically, the assumption corresponds to the idea 
that quantum states can be decomposed on a basis of ``Faraday 
lines'' excitations (as Minkowski QFT states can be decomposed on 
a particle basis).

Furthermore, this is an assumption that {\em fails\/} in conventional 
quantum field theory, because in that context well defined operators 
and finite norm states need to be smeared in at least three 
dimensions, and one-dimensional objects are too singular.\footnote{The 
assumption does not fail, however, in two-dimensional Yang-Mills 
theory, which is invariant under area preserving diffeomorphisms, and 
where loop quantization techniques were successfully employed 
\cite{ALMMT}.} The fact that at the basis of loop gravity there is a 
mathematical assumption that fails for conventional Yang-Mills quantum 
field theory is probably at the origin of some of the resistance that 
loop quantum gravity encounters among some high energy theorists.  
What distinguishes gravity from Yang-Mills theories, however, and 
makes this assumption viable in gravity even if it fails for 
Yang-Mills theory is diffeomorphism invariance.  The loop states are 
singular states that span a ``huge'' non-separable state space.  
(Non-perturbative) diffeomorphism invariance plays two roles.  First, 
it wipes away the infinite redundancy.  Second, it ``smears'' a loop 
state into a knot state, so that the physical states are not really 
concentrated in one dimension, but are, in a sense, smeared all over 
the entire manifold by the nonperturbative diffeomorphisms.  This will 
be more clear in the next section.

\subsection{Physical meaning of diffeomorphism invariance, and 
its implementation in the quantum theory} \label{relational}

Conventional field theories are not invariant under a 
diffeomorphism acting on the dynamical fields.  (Every field 
theory, suitably formulated, is trivially invariant under a 
diffeomorphism acting on {\em everything}.)  General relativity, 
on the contrary is invariant under such transformations.  More 
precisely every general relativistic theory has this property.  
Thus, diffeomorphism invariance is not a feature of just the 
gravitational field: it is a feature of physics, once the 
existence of relativistic gravity is taken into account.  Thus, 
one can say that the gravitational field is not particularly 
``special'' in this regard, but that diff-invariance is a 
property of the physical world that can be disregarded only in 
the approximation in which the dynamics of gravity is neglected.  
What is this property?  What is the physical meaning of 
diffeomorphism invariance?

Diffeomorphism invariance is the technical implementation of a 
physical idea, due to Einstein.  The idea is a deep modification 
of the pre-general-relativistic (pre-GR) notions of space and 
time.  In pre-GR physics, we assume that physical objects can be 
localized in space and time with respect to a fixed non-dynamical 
background structure.  Operationally, this background spacetime 
can be defined by means of physical reference-system objects, but 
these objects are considered as dynamically decoupled from the 
physical system that one studies.  This conceptual structure 
fails in a relativistic gravitational regime.  In general 
relativistic physics, the physical objects are localized in space 
and time only with respect to each other.  Therefore if we 
``displace'' all dynamical objects in spacetime at once, we are 
not generating a different state, but an equivalent mathematical 
description of the same physical state.  Hence, diffeomorphism 
invariance.

Accordingly, a physical state in GR is not ``located'' somewhere 
\cite{RovelliHalf,RovelliObservables,RovelliLocalization2} 
(unless an appropriate gauge fixing is made).  Pictorially, GR is 
not physics over a stage, it is the dynamical theory of (or 
including) the stage itself.

Loop quantum gravity is an attempt to implement this subtle 
relational notion of spacetime localization in quantum field 
theory.  In particular, the basic quantum field theoretical 
excitations cannot be localized somewhere (localized with respect 
to what?)\ as, say, photons are.  They are quantum excitations of 
the ``stage'' itself, not excitations over a stage.  Intuitively, 
one can understand from this discussion how knot theory plays a 
role in the theory.  First, we define quantum states that 
correspond to loop-like excitations of the gravitational field, 
but then, when factoring away diffeomorphism invariance, the 
location of the loop becomes irrelevant.  The only remaining 
information contained in the loop is then its knotting (a knot is 
a loop up to its location).  Thus, diffeomorphism invariant 
physical states are labeled by knots.  A knot represent an 
elementary quantum excitation of space.  It is not here or there, 
since it {\em is\/} the space with respect to which here and 
there can be defined.  A knot state is an elementary quantum of 
space.

In this manner, loop quantum gravity ties the new notion of space 
and time introduced by general relativity with quantum mechanics.  
As I will illustrate later on, the existence of such elementary 
quanta of space is then made concrete by the quantization of the 
spectra of geometrical quantities.

\subsection{Problems {\em not\/} addressed}

Quantum gravity is an open problem that has been investigated for 
over seventy years now.  When one contemplates two deep problems, 
one is tempted to believe that they are related.  In the history 
of physics, there are surprising examples of two deep problems 
solved by one stroke (the unification of electricity and 
magnetism and the nature of light, for instance); but also many 
examples in which a great hope to solve more than one problem at 
once was disappointed (finding the theory of strong interactions 
and getting rid of quantum field theory infinities, for 
instance).  Quantum gravity has been asked, at some time or the 
other, to address almost every deep open problem in theoretical 
physics (and beyond).  Here is a list of problems that have been 
connected to quantum gravity in the past, but about which loop 
quantum gravity has little to say:

\begin{description}
	\item[Interpretation of quantum mechanics.]  
	Loop quantum gravity is a standard quantum (field) theory.  
	Pick your favorite interpretation of quantum mechanics, and 
	use it for interpreting the quantum aspects of the theory.  I 
	will refer to two such interpretations below: When discussing 
	the quantization of area and volume, I will use the relation 
	between eigenvalues and outcomes of measurements performed 
	with classical physical apparata; when discussing evolution, I 
	will refer to the histories interpretation.  The peculiar way 
	of describing time evolution in a general relativistic theory 
	may require some appropriate variants of standard 
	interpretations, such as generalized canonical quantum theory 
	\cite{RovelliQM,RovelliTime,RovelliQM3} or Hartle's generalized 
	quantum mechanics \cite{Hartle}.  But loop quantum gravity has 
	no help to offer to the scientists that have speculated that 
	quantum gravity will solve the measurement problem.  On the 
	other hand, the spacetime formulation of loop quantum gravity 
	that has recently been developed (see Section \ref{spacetime}) 
	is naturally interpreted in terms of histories interpretations 
	\cite{Hartle,Isham,Isham2,Isham3,Isham4}.  Furthermore, I 
	think that solving the problem of the interpretation of 
	quantum mechanics might require relational ideas connected 
	with the relational nature of spacetime revealed by general 
	relativity \cite{RovelliQMint,RovelliHalf}.

	\item[Quantum cosmology.]  There is widespread confusion 
	between quantum cosmology and quantum gravity.  Quantum 
	cosmology is the theory of the entire universe as a quantum 
	system without external observer \cite{HartleQC}: with or 
	without gravity, makes no difference.  Quantum gravity is the 
	theory of one dynamical entity: the quantum gravitational 
	field (or the spacetime metric): just one field among the 
	many.  Precisely as for the theory of the quantum 
	electromagnetic field, we can always assume that we have a 
	classical observer with classical measuring apparata 
	measuring gravitational phenomena, and therefore study 
	quantum gravity disregarding quantum cosmology.  For 
	instance, the physics of a Planck size small cube is governed 
	by quantum gravity and, presumably, has no cosmological 
	implications.

	\item[Unifications of all interactions] or {\bf ``Theory of 
	Everything''}.  A common criticism to loop quantum gravity is 
	that it does not unify all interactions.  But the idea that 
	quantum gravity can be understood {\em only\/} in 
	conjunctions with other fields is an interesting hypothesis, 
	not an established truth.

	\item[Mass of the elementary particles.]  As far as I see, 
	nothing in loop quantum gravity suggests that one could compute 
	masses from quantum gravity. 

	\item[Origin of the Universe.]  It is likely that a sound 
	quantum theory of gravity will be needed to understand the 
	physics of the Big Bang.  The converse is probably not true: 
	we should be able to understand the small scale structure of 
	spacetime even if we do not understand the origin of the 
	Universe.

	\item[Arrow of time.] Roger Penrose has argued for some time 
	that it should be possible to trace the time asymmetry in the 
	observable Universe to quantum gravity. 

	\item[Physics of the mind.]  Roger has also speculated 
	that quantum gravity is responsible for the wave function 
	collapse, and, indirectly, governs the physics of the mind 
	\cite{PenroseBook}. 
		  
\end{description}

A problem that has been repeatedly tied to quantum gravity, and 
which loop quantum gravity {\em might\/} be able to address is 
the problem of the ultraviolet infinities in quantum field 
theory.  The very peculiar nonperturbative short scale structure 
of loop quantum gravity introduces a physical cutoff.  Since 
physical spacetime itself comes in quanta in the theory, there is 
literally no space in the theory for the very high momentum 
integrations that originate the ultraviolet divergences.  Lacking 
a complete and detailed calculation scheme, however, one cannot 
yet claim with confidence that the divergences, chased from the 
door, will not reenter from the window.

\section{The formalism}\label{5}  

Here, I begin the technical description of the basics of loop 
quantum gravity.  The starting point of the construction of the 
quantum theory is classical general relativity, formulated in 
terms of the Sen-Ashtekar-Barbero connection 
\cite{Sen,Ashtekar86,Barbero3}.  Detailed introductions to the 
(complex) Ashtekar formalism can be found in the book 
\cite{AshtekarBook}, in the review article \cite{Rovelli91}, or 
in the conference proceedings \cite{Ehlers94}.  The real version 
of the theory is presently the most widely used.

Classical general relativity can be formulated in phase space 
form as follows \cite{AshtekarBook,Barbero3}.  We fix a 
three-dimensional manifold $M$ (compact and without boundaries) 
and consider a smooth real $SU(2)$ connection $A_a^i(x)$ and and 
a vector density $\tilde{E}^a_i(x)$ (transforming in the vector 
representation of $SU(2)$) on~$M$.  We use $a,b,\ldots=1,2,3$ for 
spatial indices and $i,j,\ldots=1,2,3$ for internal indices.  The 
internal indices can be viewed as labeling a basis in the Lie 
algebra of $SU(2)$ or the three axis of a local triad.  We 
indicate coordinates on $M$ with~$x$.  The relation between these 
fields and conventional metric gravitational variables is as 
follows: $\tilde{E}^a_i(x)$ is the (densitized) inverse triad, 
related to the three-dimensional metric $g_{ab}(x)$ of 
constant-time surfaces by 
 \begin{equation} 
 g\ g^{ab} = \tilde{E}^a_i\tilde{E}^b_i, 
 \end{equation} 
 where $g$ is the determinant of $g_{ab}$; and 
\begin{equation} 
A_a^i(x)=\Gamma_a^i(x)+ \gamma\ k_a^i(x); 
\label{real} 
\end{equation} 
$\Gamma_a^i(x)$ is the spin connection associated to the triad, 
(defined by $\partial_{[a}e_{b]}^{i}=\Gamma_{[a}^i e_{b]j}$, where 
$e_{a}^{i}$ is the triad).  $k_a^i(x)$ is the extrinsic curvature 
of the constant time three surface.  

In (\ref{real}), $\gamma$ is a constant, denoted the Immirzi 
parameter, that can be chosen arbitrarily (it will enter the 
hamiltonian constraint) \cite{Immirzi2,Immirzi3,Immirzi}.  
Different choices for $\gamma$ yield different versions of the 
formalism, all equivalent in the classical domain.  If we choose 
$\gamma$ to be equal to the imaginary unit, $\gamma=\sqrt{-1}$, 
then $A$ is the standard Ashtekar connection, which can be shown  
to be the projection of the selfdual part of the four-dimensional 
spin connection on the constant time surface.  If we choose 
$\gamma=1$, we obtain the real Barbero connection.  The 
hamiltonian constraint of Lorentzian general relativity has a 
particularly simple form in the $\gamma=\sqrt{-1}$ formalism, 
while the hamiltonian constraint of Euclidean general relativity 
has a simple form when expressed in terms of the $\gamma=1$ real 
connection.  Other choices of $\gamma$ are viable as well.  In 
particular, it has been argued that the quantum theory based on 
different choices of $\gamma$ are genuinely physical 
inequivalent, because they yield ``geometrical quanta'' of 
different magnitude \cite{RovelliThiemann}.  Apparently, there is 
a unique choice of $\gamma$ yielding the correct $1/4$ 
coefficient in the Bekenstein-Hawking formula 
\cite{Krasnov,Krasnov2,Rovelli96,AshtekarEtAl97,%
RovelliAscona,CorichiKrasnov}, 
but the matter is still under discussion.

The spinorial version of the Ashtekar variables is given in terms 
of the Pauli matrices $\sigma_i, i=1,2,3$, or the $su(2)$ 
generators ${\tau_i} = - \frac{{\rm i}}{2} \ {\sigma_i}$, by 
\begin{eqnarray}
\tilde{E}^{a}(x) 
        &=& - {\rm i}\ \tilde{E}^a_i(x) \, {\sigma_i}
         =  2 \tilde{E}^a_i(x) \, {\tau_i},  \\
A_{a}(x)&=& - \frac{{\rm i}}{{2}}\ {A}_a^i(x)\, {\sigma_i}
         = {A}_a^i(x)\, {\tau_i}\ \ .
\end{eqnarray}
Thus, $A_a(x)$ and $\tilde{E}^a(x)$ are $2\times 2$ 
anti-hermitian complex matrices. 

The theory is invariant under local $SU(2)$ gauge 
transformations, three-dimensional diffeomorphisms of the 
manifold on which the fields are defined, as well as under 
(coordinate) time translations generated by the hamiltonian 
constraint.  The full dynamical content of general relativity is 
captured by the three constraints that generate these gauge 
invariances \cite{Sen,AshtekarBook}.  

As already mentioned, the Lorentzian hamiltonian constraint does 
not have a simple polynomial form if we use the real connection 
(\ref{real}).  For a while, this fact was considered an obstacle 
for defining the quantum hamiltonian constraint; therefore the 
complex version of the connection was mostly used.  However, 
Thiemann has recently succeeded in constructing a Lorentzian 
quantum hamiltonian constraint 
\cite{Thiemann96,Thiemann96b,Thiemann96c} in spite of the 
non-polynomiality of the classical expression.  This is the 
reason why the real connection is now widely used.  This choice 
has the advantage of eliminating the old ``reality conditions'' 
problem, namely the problem of implementing non-trivial reality 
conditions in the quantum theory.

\subsection{Loop algebra} \label{LoopAlgebra}

Certain classical quantities play a very important role in the 
quantum theory.  These are: the trace of the holonomy of the 
connection, which is labeled by loops on the three manifold; and 
the higher order loop variables, obtained by inserting the $E$ 
field (in $n$ distinct points, or ``hands'') into the holonomy 
trace.  More precisely, given a loop $\alpha$ in $M$ and the 
points $s_1,s_2,\ldots,s_n\in\alpha$ we define:
 \begin{eqnarray}
  {\cal T}[\alpha]          &=& - {\rm Tr} [U_\alpha] ,
\label{t0}  \\
  {\cal T}^a[\alpha](s)     
      &=& - {\rm Tr} [U_\alpha(s,s) \tilde{E}^a(s)] 
\label{t1}
\end{eqnarray}
and, in general
\begin{eqnarray}
&&{\cal T}^{a_1a_2}[\alpha](s_1,s_2) = \nonumber \\  
&&\ \ =  - {\rm Tr} [U_\alpha(s_1,s_2) \tilde{E}^{a_2}(s_2) 
                U_\alpha(s_2,s_1) \tilde{E}^{a_1}(s_1) ] ,
\label{tn} \\
&& {\cal T}^{a_1\ldots a_N}[\alpha](s_1 \ldots s_N) = 
\nonumber \\  
&& \ \ = 
- {\rm Tr} [U_\alpha(s_1,s_N) \tilde{E}^{a_N}(s_N)
                U_\alpha(s_N,s_{N-1}) \ldots 
\tilde{E}^{a_1}(s_1) ]  
\nonumber
\end{eqnarray}
 where $U_{\alpha}(s_{1},s_{2})\sim{\cal P} exp 
 \{\int_{s_{1}}^{s_{2}} 
 A_{a}(\alpha(s))ds \}$ is the parallel propagator of $A_a$ along 
 $\alpha$, defined by 
 \begin{equation}
	 \frac{d}{ds}U_{\alpha}(1,s) = \frac{d\alpha_{a}(s)}{ds} \ 
	 A_{a}(\alpha(s))\ U_{\alpha}(1,s)
 \end{equation}
  (See \cite{DePietriRovelli} for more details.)  These are the 
  loop observables, introduced in Yang Mills theories in 
  \cite{GambiniTrias2,GambiniTrias}, and in gravity in 
  \cite{RovelliSmolin88,RovelliSmolin90}.

  The loop observables coordinatize the phase space and have a 
  closed Poisson algebra, denoted the loop algebra.  This algebra 
  has a remarkable geometrical flavor.  For instance, the Poisson 
  bracket between ${\cal T}[\alpha]$ and ${\cal T}^a[\beta](s)$ 
  is non vanishing only if $\beta(s)$ lies over $\alpha$; if it 
  does, the result is proportional to the holonomy of the Wilson 
  loops obtained by joining $\alpha$ and $\beta$ at their 
  intersection (by rerouting the 4 legs at the intersection).  
  More precisely
\begin{equation}
	\{{\cal T}[\alpha], {\cal T}^a[\beta](s) \} =
	\Delta^{a}[\alpha,\beta(s)] \ \left[{\cal T}
	[\alpha\#\beta]-{\cal T}[\alpha\#\beta^{-1}]\right].
\end{equation}
Here 
\begin{equation}
		\Delta^{a}[\alpha,x] = \int ds\ 
		\frac{d\alpha^{a}(s)}{ds}\ \delta^{3}(\alpha(s),x)
	\label{Delta}
\end{equation}
is a vector distribution with support on $\alpha$ and 
$\alpha\#\beta$ is the loop obtained starting at the intersection 
between $\alpha$ and $\beta$, and following first $\alpha$ and 
then $\beta$.  $\beta^{-1}$ is $\beta$ with reversed orientation.

A (non-SU(2) gauge invariant) quantity that plays a role in 
certain aspects of the theory, particularly in the regularization 
of certain operators, is obtained by integrating the $E$ field 
over a two dimensional surface $S$ 
\begin{equation}
	E[S,f] = \int_{S} dS_{a}\ \tilde{E}^{a}_{i}\ f^{i}, 
\end{equation}
where $f$ is a function on the surface $S$, taking values in the 
Lie algebra of $SU(2)$.  In alternative to the full loop 
observables (\ref{t0},\ref{t1},\ref{tn}), one also can take the 
holonomies and $E[S,f]$ as elementary variables 
\cite{AshtekarLewandowski3,AshtekarLewandowskiArea2}; this is 
more natural to do, for instance, in the C*-algebric approach 
\cite{AshtekarIsham}.

\subsection{Loop quantum gravity} \label{hilbertspace}

The kinematic of a quantum theory is defined by an algebra of 
``elementary'' operators (such as $x$ and $i\hbar d/dx$, or 
creation and annihilation operators) on a Hilbert space $\cal H$.  
The physical interpretation of the theory is based on the 
connection between these operators and classical variables, and 
on the interpretation of $\cal H$ as the space of the quantum 
states.  The dynamics is governed by a hamiltonian, or, as in 
general relativity, by a set of quantum constraints, constructed 
in terms of the elementary operators.  To assure that the quantum 
Heisenberg equations have the correct classical limit, the 
algebra of the elementary operator has to be isomorphic to the 
Poisson algebra of the elementary observables.  This yields the 
heuristic quantization rule: ``promote Poisson brackets to 
commutators''.  In other words, define the quantum theory as a 
linear representation of the Poisson algebra formed by the 
elementary observables.  For the reasons illustrated in section 
\ref{4}, the algebra of elementary observables we choose for the 
quantization is the loop algebra, defined in section 
\ref{LoopAlgebra}.  Thus, the kinematic of the quantum theory is 
defined by a unitary representation of the loop algebra.  Here, I 
construct such representation following a simple path.

We can start ``\`{a} la Schr\"odinger'' by expressing quantum 
states by means of the amplitude of the connection, namely by 
means of functionals $\Psi(A)$ of the (smooth) connection.  These 
functionals form a linear space, which we promote to a Hilbert 
space by defining a inner product.  To define the inner product, 
we choose a particular set of states, which we denote 
``cylindrical states'' and begin by defining the 
scalar product between these.   

Pick a graph $\Gamma$, say with $n$ links, denoted 
$\gamma_{1}\ldots\gamma_{n}$, immersed in the manifold $M$.  For 
technical reasons, we require the links to be analytic.  Let 
$U_{i}(A)=U_{\gamma_{i}}, \ i=1,\ldots, n$ be the parallel transport 
operator of the connection $A$ along $\gamma_{i}$.  $U_{i}(A)$ is an 
element of $SU(2)$.  Pick a function $f(g_{1}\ldots g_{n})$ on 
$[SU(2)]^{n}$.  The graph $\Gamma$ and the function $f$ determine a 
functional of the connection as follows
\begin{equation}
	\psi_{\Gamma,f}(A)=f(U_{1}(A),\ldots, U_{n}(A)). 
\end{equation}
(These states are called cylindrical states because they were 
introduced in 
\cite{AshtekarLewandowski,AshtekarLewandowski3,%
AshtekarLewandowski2} as cylindrical functions 
for the definition of a cylindrical measure.)  Notice that we can 
always ``enlarge the graph'', in the sense that if $\Gamma$ is a 
subgraph of $\Gamma'$ we can always write
\begin{equation}
	\psi_{\Gamma,f}(A)=	\psi_{\Gamma',f'}(A). 
	\label{prmap}
\end{equation}
by simply choosing $f'$ independent from the $U_{i}$'s of the 
links which are in $\Gamma'$ but not in $\Gamma$.  Thus, given 
any two cylindrical functions, we can always view them as having 
the same graph (formed by the union of the two graphs).  Given this 
observation, we define the scalar product between any two 
cylindrical functions
 \cite{JerzyGraph,AshtekarLewandowski,AshtekarLewandowski3,%
 AshtekarLewandowski2} by
\begin{equation}
	(\psi_{\Gamma,f}, \psi_{\Gamma,h}) = \int _{SU(2)^{n}} 
	\!\!\! dg_{1}\ldots dg_{n} \  
	\overline{f(g_{1}\ldots g_{n})}\, 
	h(g_{1}\ldots g_{n}).
\label{scalarpr}
\end{equation}
where $dg$ is the Haar measure on $SU(2)$.  This scalar product 
extends by linearity to finite linear combinations of cylindrical 
functions.  It is not difficult to show that (\ref{scalarpr}) 
defines a well defined scalar product on the space of these 
linear combinations.  Completing the space of these linear 
combinations in the Hilbert norm, we obtain a Hilbert space $\cal 
H$.  This is the (unconstrained) quantum state space of loop 
gravity.\footnote{This construction of $\cal H$ as the closure 
of the space of the cylindrical functions of smooth connections
in the scalar product (\ref{scalarpr}) shows that $\cal H$ can
be defined without the need of recurring to $C^{*}$ algebraic 
techniques, distributional connections or the Ashtekar-Lewandowski 
measure. The casual reader, however, should be warned that 
the resulting $\cal H$ topology is different than the natural 
topology on the space of connections: if a sequence $\Gamma_{n}$ of 
graphs converges pointwise to a graph $\Gamma$, the corresponding
cylindrical functions $\psi_{\Gamma_{n},f}$ do not converge to 
$\psi_{\Gamma,f}$ in the $\cal H$ Hilbert space topology.}
$\cal H$ carries a natural unitary representation of the 
diffeomorphism group and of the group of the local $SU(2)$ 
transformations, obtained transforming the argument of the 
functionals.  An important property of the scalar product 
(\ref{scalarpr}) is that it is invariant under both these 
transformations.

$\cal H$ is non-separable.  At first sight, this may seem as a 
serious obstacle for its physical interpretation.  But we will 
see below that after factoring away diffeomorphism invariance we 
may obtain a separable Hilbert space (see section 
\ref{diffinvariance}).  Also, standard spectral theory holds on 
$\cal H$, and it turns out that using spin networks (discussed 
below) one can express $\cal H$ as a direct sum over finite 
dimensional subspaces which have the structure of Hilbert spaces 
of spin systems; this makes practical calculations very 
manageable.

Finally, we will use a Dirac notation and write 
\begin{equation}
	\Psi(A) = \langle A | \Psi \rangle,
	\label{dirac}
\end{equation}
in the same manner in which one may write $\psi(x) = \langle x | 
\Psi \rangle$ in ordinary quantum mechanics. As in that case, 
however, we should remember that $|A\rangle$ is not a 
normalizable state. 

\subsection{Loop states and spin network states}
\label{states}

A subspace ${\cal H}_{0}$ of $\cal H$ is formed by states 
invariant under $SU(2)$ gauge transformations.  We now define an 
orthonormal basis in ${\cal H}_{0}$.  This basis represents a 
very important tool for using the theory.  It was introduced in 
\cite{RovelliSmolin95b} and developed in \cite{Baez95a,Baez95aa}; 
it is denoted spin network basis.

First, given a loop $\alpha$ in $M$, there is a normalized state 
$\psi_{\alpha}(A)$ in $\cal H$, which is obtained by taking 
$\Gamma=\alpha$ and $f(g)=-Tr(g)$.  Namely
\begin{equation}
	\psi_{\alpha}(A) = - {\rm Tr} U_{\alpha}(A). 
\end{equation}
We introduce a Dirac notation for the abstract states, and 
denote this state as $|\alpha\rangle$.  These sates are called 
loop states. Using Dirac notation, we can write
\begin{equation}
	\psi_{\alpha}(A) = \langle A |\alpha\rangle,
\end{equation}
It is easy to show that loop states are normalizable.  Products 
of loop states are normalizable as well.  Following tradition, we 
denote with $\alpha$ also a multiloop, namely a collection of 
(possibly overlapping) loops $\{\alpha_{1},\ldots,\alpha_{n},\}$, 
and we call 
\begin{equation}
	\psi_{\alpha}(A)=\psi_{\alpha_{1}}(A)\times \ldots \times 	
	\psi_{\alpha_{n}}(A)
\end{equation}
a multiloop state.  (Multi-)loop states represented the main tool 
for loop quantum gravity before the discovery of the spin network 
basis.  Linear combinations of multiloop states (over-)span $\cal 
H$, and therefore a generic state $\psi(A)$ is fully 
characterized by its projections on the multiloop states, namely 
by
\begin{equation}
	\psi(\alpha) = (\psi_{\alpha},\psi).
	\label{looptransform1}
\end{equation}
The ``old'' loop representation was based on representing quantum 
states in this manner, namely by means of the functionals 
$\psi(\alpha)$ over loop space defined in(\ref{looptransform1}).  
Equation (\ref{looptransform1}) can be explicitly written as an 
integral transform, as we will see in section \ref{otherstructures}. 

Next, consider a graph $\Gamma$. A ``coloring'' of $\Gamma$ is 
given by the following. 
\begin{enumerate}
 \item Associate an irreducible representation of $SU(2)$ to each 
 link of $\Gamma$.  Equivalently, we may associate to each link 
 $\gamma_{i}$ a half integer number $s_{i}$, the spin of the 
 irreducible, or, equivalently, an integer number $p_{i}$, the 
 ``color'' $p_{i} = 2s_{i}$.  
 
 \item Associate an invariant tensor $v$ in the tensor product of 
 the representations $s_{1}\ldots s_{n}$, to each node of $\Gamma$ 
 in which links with spins $s_{1}\ldots s_{n}$ meet.  An invariant 
 tensor is an object with $n$ indices in the representations 
 $s_{1}\ldots s_{n}$ that transform covariantly.  If $n=3$, there 
 is only one invariant tensor (up to a multiplicative factor), 
 given by the Clebsh-Gordon coefficient.  An invariant tensor is 
 also called an {\it intertwining tensor}.  All invariant tensors 
 are given by the standard Clebsch-Gordon theory.  More precisely, 
 for fixed $s_{1}\ldots s_{n}$, the invariant tensors form a 
 finite dimensional linear space.  Pick a basis $v_{j}$ is this 
 space, and associate one of these basis elements to the node.  
 Notice that invariant tensors exist only if the tensor product of 
 the representations $s_{1}\ldots s_{n}$ contains the trivial 
 representation.  This yields a condition on the coloring of the 
 links.  For $n=3$, this is given by the well known Clebsh-Gordan 
 condition: each color is not larger than the sum of the other 
 two, and the sum of the three colors is even.
\end{enumerate}
We indicate a colored graph by $\{\Gamma, \vec s, \vec v\}$, or 
simply $S=\{\Gamma, \vec s, \vec v\}$, and denote it a ``spin 
network''.  (It was Penrose who first had the intuition that this 
mathematics could be relevant for describing the quantum 
properties of the geometry, and who gave the first version of 
spin network theory \cite{Penrose2,Penrose}.)

Given a spin network $S$, we can construct a state $\Psi_{S}(A)$ 
as follows.  We take the propagator of the connection along each 
link of the graph, in the representation associated to that link, 
and then, at each node, we contract the matrices of the 
representation with the invariant tensor.  We obtain a state 
$\Psi_{S}(A)$, which we also write as
\begin{equation}
	\psi_{S}(A) = \langle A |S \rangle . 
\end{equation}
One can then show the following. 
\begin{itemize}
 \item The spin network states are normalizable.  The 
 normalization factor is computed in \cite{DePietriRovelli}.  
 \item They are $SU(2)$ gauge invariant.  
 \item Each spin network state can be decomposed into a finite 
 linear combination of products of loop states.
 \item The (normalized) spin network states form an orthonormal 
 basis for the gauge $SU(2)$ invariant states in $\cal H$ 
 (choosing the basis of invariant tensors appropriately).
 \item The scalar product between two spin network states can be 
 easily computed graphically and algebraically.  See 
 \cite{DePietriRovelli} for details.
\end{itemize}
The spin network states provide a very convenient basis for the 
quantum theory.  

The spin network states defined above are $SU(2)$ gauge 
invariant.  There exists also an extension of the spin network 
basis to the full Hilbert space (see for instance 
\cite{AshtekarLewandowskiArea2,BorissovEtAl97}, and references 
therein).

\subsection{Relation between spin network states and loop 
states and diagrammatic representation of the states}

A diagrammatic representation for the states in $\cal H$ is very 
useful in concrete calculations. 
First, associate to a loop state $|\alpha\rangle$ a diagram 
in $M$, formed by the loop $\alpha$ itself.  Next, notice that we 
can {\em multiply\/} two loop states, obtaining a normalizable 
state.  We represent the product of $n$ loop states by the 
diagram formed by the set of the $n$ (possibly overlapping) 
corresponding loops (we denote this set ``multiloop'').  Thus, 
linear combinations of multiloops diagrams represent states in 
$\cal H$.  Representing states as linear combinations of 
multiloops diagrams makes computation in $\cal H$  
easy.

Now, the spin network state defined by the graph with no nodes 
$\alpha$, with color 1, is clearly, by definition, the loop state 
$|\alpha\rangle$, and we represent it by the diagram $\alpha$.  
The spin network state $|\alpha, n\rangle$ determined by the graph 
without nodes $\alpha$, with color $n$ can be obtained as 
follows.  Draw $n$ parallel lines along the loop $\alpha$; cut 
all lines at an arbitrary point of $\alpha$, and consider the 
$n!$ diagrams obtained by joining the legs after a permutation.  
The linear combination of these $n!$ diagrams, taken with 
alternate signs (namely with the sign determined by the parity of 
the permutation) is precisely the state $|\alpha, n\rangle$.  The 
reason of this key result can be found in the fact that an  
irreducible representation of $SU(2)$ can be obtained as the 
totally symmetric tensor product of the fundamental 
representation with itself.  For details, see 
\cite{DePietriRovelli}.

\begin{figure}
$$
\begin{array}{c}\mbox{\epsfig{file=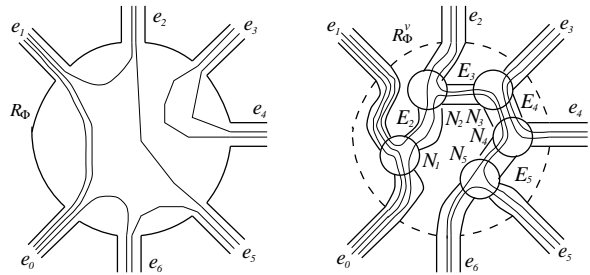}}\end{array}
$$
\caption{Construction of ``virtual'' nodes and ``virtual''
links over an $n$-valent node.}
\label{TheVirtualNode}
\end{figure}

Next, consider a graph $\Gamma$ with nodes.  Draw $n_{i}$ parallel 
lines along each link $\gamma_{i}$.  Join pairwise the end points 
of these lines at each node (in an arbitrary manner), in such a 
way that each line is joined with a line from a different link 
(see Figure \ref{TheVirtualNode}).  In this manner, one obtain a 
multiloop diagram.  Now antisymmetrize the $n_{i}$ parallel lines 
along each link, obtaining a linear combination of diagrams 
representing a state in $\cal H$.  One can show that this state is 
a spin network state, where $n_{i}$ is the color of the links and 
the color of the nodes is determined by the pairwise joining of 
the legs chosen \cite{DePietriRovelli}.  Again, simple $SU(2)$ 
representation theory is behind this result.

More in detail, if a node is trivalent (has 3 adjacent links), 
then we can join legs pairwise only if the total number of the legs 
is even, and if the number of the legs in each link is smaller 
or equal than the sum of the number of the other two. This can 
be immediately recognized as the Clebsch-Gordan condition. If 
these conditions are satisfied, there is only a single way of 
joining legs. This corresponds to the fact that there is only one 
invariant tensor in the product of three irreducible of $SU(2)$. 
Higher valence nodes can be decomposed in trivalent ``virtual'' 
nodes, joined by ``virtual'' links.  Orthogonal independent 
invariant tensors are obtained by varying over all allowed 
colorings of these virtual links (compatible with the 
Clebsch-Gordan conditions at the virtual nodes). Different 
decompositions of the node give different orthogonal bases. 
Thus the total (links and nodes) coloring of a spin network can 
be represented by means of the coloring of the real and the 
virtual links. See Figure \ref{TheVirtualNode}. 

Viceversa, multiloop states can be decomposed in spin network states 
by simply symmetrizing along (real and virtual) nodes.  This can be 
done particularly easily diagrammatically, as illustrated by the 
graphical formulae in \cite{RovelliSmolin95b,DePietriRovelli}.  These 
are standard formulae.  In fact, it is well known that the tensor 
algebra of the $SU(2)$ irreducible representations admits a completely 
graphical notation.  This graphical notation has been widely used for 
instance in nuclear and atomic physics.  One can find it presented in 
detail in books such as 
\cite{GraphMethods,GraphMethods2,GraphMethods3}.  The application of 
this diagrammatic calculus to quantum gravity is described in detail 
in \cite{DePietriRovelli}, which I recommend to anybody who intends to 
perform concrete calculations in loop gravity.

It is interesting to notice that loop quantum gravity was first 
constructed in a pure diagrammatic notation, in 
\cite{RovelliSmolin90}.  The graphical nature of this calculus 
puzzled some, and the theory was accused of being vague and 
strange.  Only later the diagrammatic notation was recognized to 
be the very conventional $SU(2)$ diagrammatic calculus, well 
known in atomic and nuclear physics.\footnote{The following 
anecdote illustrates the level of the initial confusion.  The 
first elaborate graphical computation of the eigenvalues of the 
area led to the mysterious formula $A_{p}=\frac{1}{2}\hbar G 
\sqrt{p^{2}+2p}$, with integer $p$.  One day, Junichi Iwasaki, at 
the time a student, stormed into my office and told me to rewrite 
the formula in terms of the half integer $j=\frac{1}{2}p$.  This 
yields the very familiar expression for the 
eigenvalues of the $SU(2)$ Casimir $A_{j}=\hbar G \sqrt{j(j+1)}$.  
Nobody had previously noticed this fact.}

\subsection{The representation}

I now define the quantum operators, corresponding to the
$\cal T$-variables, as linear operators on $\cal H$. These
form a representation of the loop variables Poisson algebra.
The operator ${\cal T}[\alpha]$ acts diagonally 
\begin{eqnarray}
 {\cal T}[\alpha] \Psi(A)  = - \mathrm{Tr}\, U_{\alpha}(A)\ \Psi(A).
\end{eqnarray} 
(Recall that products of loop states and spin network states are
normalizable states). In diagrammatic notation, the operator 
simply adds a loop to a (linear combination of) multiloops
\begin{eqnarray}
 {\cal T}[\alpha]\ |\Psi\rangle  = |\alpha\rangle |\Psi\rangle. 
 \label{t00}
\end{eqnarray} 

Higher order loop operators are expressed in terms of the 
elementary ``grasp'' operation.  Consider first the operator 
${\cal T}^{a}(s)[\alpha]$, with one hand in the point 
$\alpha(s)$. The operator annihilates all loop states that do 
not cross the point $\alpha(s)$. Acting on a loop state 
$|\beta\rangle$, it gives
\begin{equation}
	{\cal T}^{a}(s)[\alpha]\ |\beta\rangle=
	 l^2_0 \ \Delta^a[\beta,\alpha(s)] \ 
     \left[ |\alpha\#\beta\rangle - |\alpha\#\beta^{-1}\rangle 
     \right],
\label{t11}
\end{equation}
where we have introduced the elementary length $l_0$ by
\begin{equation}
 l^2_0 = \hbar G = \frac{16 \pi \hbar G_{{\rm Newton}}}{c^3} 
 = 16 \pi\ l^2_{Planck}
\end{equation}
and $\Delta^a$ and $\#$ are defined in section \ref{LoopAlgebra}.  
This action extends by linearity, continuity and by the Leibniz 
rule to products and linear combinations of loop states, and to 
the full $\cal H$.  In particular, it is not difficult to compute 
its action on a spin network state \cite{DePietriRovelli}.  
Higher order loop operators act similarly.  It is a simple 
exercise to verify that these operators provide a representation 
of the classical Poisson loop algebra.

All the operators in the theory are then constructed in terms of 
these basics loop operators, in the same way in which in 
conventional QFT one constructs all operators, including the 
hamiltonian, in terms of creation and annihilation operators.  
The construction of the composite operators requires the 
development of regularization techniques that can be used in the 
absence of a background metric.  These have been introduced in 
\cite{Smolin93} and developed in 
\cite{RovelliSmolin95,DePietriRovelli,AshtekarEtAl95,%
weave,LewandowskiArea,AshtekarLewandowskiArea2}.

\subsection{Algebraic version (``loop representation'') and 
differential version (``connection representation'')  of the 
formalism, and their equivalence}

Imagine we want to quantize the one dimensional harmonic oscillator.  
We can consider the Hilbert space of square integrable functions 
$\psi(x)$ on the real line, and express the momentum and the 
hamiltonian as differential operators.  Denote the 
eigenstates of the hamiltonian as $\psi_{n}(x)=\langle x| n\rangle$.  
It is well known that the theory can be expressed entirely in 
algebraic form in terms of the states $|n\rangle$.  In doing so, all 
elementary operators are algebraic: $ \hat x |n\rangle = 
\frac{1}{\sqrt{2}}(|n-1\rangle + (n+1) |n+1\rangle)$, $ \hat p 
|n\rangle = \frac{-i}{\sqrt{2}}(|n-1\rangle - (n+1) |n+1\rangle)$.  
Similarly, in quantum gravity we can directly construct the quantum 
theory in the spin-network (or loop) basis, without ever mentioning 
functionals of the connections.  This representation of the theory is 
denoted the ``loop representation''.

A section of the first paper on loop quantum gravity by Rovelli 
and Smolin \cite{RovelliSmolin90} was devoted to a detailed 
study of ``transformation theory'' (in the sense of Dirac) on the 
state space of quantum gravity, and in particular on the 
relations between the loop states
\begin{equation}
           \psi(\alpha)=\langle\alpha|\psi\rangle
\end{equation}
and the states $\psi(A)$ giving the amplitude for a connection field 
configuration $A$, and defined by
\begin{equation}
           \psi(A)=\langle A|\psi\rangle. 
\end{equation}
Here $|A\rangle$ are ``eigenstates of the connection operator'', 
or, more precisely (since the operator corresponding to the 
connection is ill defined in the theory) the generalized states 
that satisfy
\begin{equation}
    T[\alpha]\ |A\rangle= - Tr[{\cal P}e^{\int_{\alpha}A}]\ |A\rangle. 
\end{equation}
However, at the time of \cite{RovelliSmolin90} the lack of a scalar 
product made transformation theory quite involved. 

On the other hand, the introduction of the scalar product 
(\ref{scalarpr}) gives a rigorous meaning to the loop 
transform.  In fact, we can 
write, for every spin network $S$, and every state $\psi(A)$
  \begin{equation}
	\psi(S)=\langle S|\psi\rangle=
	(\psi_{S},\psi). 
  \end{equation}
This equation defines a unitary mapping between the two 
presentations of $\cal H$: the ``loop representation'', in 
which one works in terms of the basis $|S\rangle$; and the 
``connection representation'', in which one uses wave 
functionals $\psi(A)$.

The development of the connection representation followed a 
winding path through $C^{*}$-algebraic \cite{AshtekarIsham} and 
measure theoretical 
\cite{AshtekarLewandowski,AshtekarLewandowski2,%
AshtekarLewandowski3} methods.  The work of 
Ashtekar, Isham, Lewandowski, Marolf, Mourao and Thiemann has 
finally put the connection representation on a firm ground, and, 
indirectly has much clarified the mathematics underlying the 
original loop approach.  In the course of this development, doubts 
were raised about the precise relations between the connection and 
the loop formalisms.  Today, the complete equivalence of these two 
approaches (always suspected) has been firmly established.  In 
particular, the work of Roberto DePietri \cite{DePietri} has 
proven the unitary equivalence of the two formalisms.  For a 
recent discussion see also \cite{Lewandowski97}.

\subsection{Other structures in $\cal H$} \label{otherstructures}

The recent developments in the mathematical foundations of the 
connection representation have increased the mathematical rigor 
of the theory, raising it to the standards of mathematical 
physics.  This has been obtained at the price of introducing 
heavy mathematical tools, often unfamiliar to the average 
physicist, perhaps widening the language gap between the quantum 
gravity and the high energy physics community. 

The reason for searching a mathematical-physics level of precision is 
that in quantum gravity one moves on a very unfamiliar terrain 
--quantum field theory on manifolds-- where the experience accumulated 
in conventional quantum field theory is often useless and sometimes 
even misleading.  Given the unlikelihood of finding direct 
experimental corroboration, the research can only aim, at least for 
the moment, at the goal of finding a {\em consistent\/} theory, with 
the correct limits in the regimes that we control experimentally.  In 
these conditions, high mathematical rigor is the only assurance of the 
consistency of the theory.  In the development of quantum field theory 
mathematical rigor could be very low because extremely accurate 
empirical verifications assured the physicists that ``the theory may 
be mathematically meaningless, but it is nevertheless physical 
correct, and therefore the theory must make sense even if we do not 
understand how.''  Here, such an indirect experimental reassurance is 
lacking and the claim that the theory is well founded can be based 
only on a solid mathematical control of the theory.

One may object that a rigorous definition of quantum gravity is a 
vain hope, given that we do not even have a rigorous definition of 
QED, presumably a much simpler theory.  The objection is 
particularly valid from the point of view of a  
physicist who views gravity ``just as any other field theory 
like the ones we already understand''.  But the (serious) 
difficulties of QED and of the other conventional field theories 
are ultraviolet.  The physical hope supporting the quantum 
gravity research program is that the ultraviolet structure of a 
diffeomorphism invariant quantum field theory is profoundly 
different from the one of conventional theories.  Indeed, recall 
that in a very precise sense there is no short distance limit in 
the theory; the theory naturally cuts itself off at the Planck 
scale, due to the very quantum discreteness of spacetime.  Thus 
the hope that quantum gravity could be defined rigorously may be 
optimistic, but it is not ill founded.

After these comments, let me briefly mention some of the 
structures that have been explored in $\cal H$. 
First of all, the spin network states satisfy the Kauffman axioms 
of the tangle theoretical version of recoupling theory 
\cite{Kauffman94} (in the ``classical'' case $A=-1$) at all the 
points (in 3d space) where they meet.  (This fact is 
often misunderstood: recoupling theory lives in 2d and is 
associated by Kauffman to knot theory by means of the usual 
projection of knots from 3d to 2d.  Here, the Kauffman axioms 
are not satisfied at the intersections created by the 2d 
projection of the spin network, but only at the nodes 
in 3d.  See \cite{DePietriRovelli} for a detailed discussion.)  
For instance, consider a 4-valent node of four links 
colored $a, b, c, d$.  The color of the node is determined by 
expanding the 4-valent node into a trivalent tree; in 
this case, we have a single internal links.  The expansion can be 
done in different ways (by pairing links differently).  These are 
related to each other by the recoupling theorem of pg.\ 60 in 
Ref.\ \cite{Kauffman94}
\begin{equation}
\begin{array}{c}\setlength{\unitlength}{1 pt}
\begin{picture}(50,50)
          \put( 0,0){$a$}\put( 0,30){$b$}
          \put(45,0){$d$}\put(45,30){$c$}
          \put(10,10){\line(1,1){10}}\put(10,30){\line(1,-1){10}}
          \put(30,20){\line(1,1){10}}\put(30,20){\line(1,-1){10}}
           \put(20,20){\line(1,0){10}}\put(22,25){$j$}
          \put(20,20){\circle*{3}}\put(30,20){\circle*{3}}
\end{picture}\end{array}
    = \sum_i  \left\{\begin{array}{ccc}
                      a  & b & i \\
                      c  & d & j
              \end{array}\right\}
\begin{array}{c}\setlength{\unitlength}{1 pt}
\begin{picture}(40,40)
      \put( 0,0){$a$}\put( 0,40){$b$}
      \put(35,0){$d$}\put(35,40){$c$}
     \put(10,10){\line(1,1){10}}\put(10,40){\line(1,-1){10}}
      \put(20,30){\line(1,1){10}}\put(20,20){\line(1,-1){10}}
      \put(20,20){\line(0,1){10}}\put(22,22){$i$}
      \put(20,20){\circle*{3}}\put(20,30){\circle*{3}}
\end{picture}\end{array}
\label{rec}
\end{equation}
where the quantities 
$\left\{\begin{array}{ccc}
a & b & i \\ c & d & j  \end{array}\right\}$ 
are $su(2)$
six-j symbols (normalized as in \cite{Kauffman94}). Equation 
(\ref{rec}) follows just from the definitions given above. Recoupling 
theory provides a powerful computational tool in this context. 

Since spin network states satisfy recoupling theory, they form a 
Temperley-Lieb algebra \cite{Kauffman94}.  The scalar product 
(\ref{scalarpr}) in $\cal H$ is given also by the Temperley-Lieb 
trace of the spin networks, or, equivalently by the Kauffman 
brackets, or, equivalently, by the chromatic evaluation of the 
spin network.  See Ref.~\cite{DePietriRovelli} for an extensive 
discussion of these relations. 

Next, $\cal H$ admits a rigorous representation as an $L_{2}$ 
space, namely a space of square integrable functions.  To obtain 
this representation, however, we have to extend the notion of 
connection, to a notion of ``distributional connection''.  The 
space of the distributional connections is the closure of the 
space of smooth connection in a certain topology.  Thus, 
distributional connections can be seen as limits of sequences of 
connections, in the same manner in which distributions can be 
seen as limits of sequences of functions.  Usual distributions 
are defined as elements of the topological dual of certain spaces 
of functions.  Here, there is no natural linear structure in the 
space of the connections, but there is a natural duality between 
connections and curves in $M$: a smooth connection $A$ assigns a 
group element $U_{\gamma}(A)$ to every segment $\gamma$.  The 
group elements satisfy certain properties.  For instance if  
$\gamma$ is the composition of the two segments $\gamma_{1}$ and 
$\gamma_{2}$, then $U_{\gamma}(A) = U_{\gamma_{1}}(A) 
U_{\gamma_{2}}(A)$. 

A generalized connection $\bar{A}$ is defined as a map that 
assigns an element of $SU(2)$, which we denote as 
$U_{\gamma}(\bar{A})$ or $\bar{A}(\gamma)$ to each (oriented) 
curve $\gamma$ in $M$, satisfying the following requirements: i) 
$\bar{A}(\gamma^{-1}) = (\bar{A}(\gamma))^{-1}$; and, ii) 
$\bar{A}(\gamma_2\circ \gamma_1) = \bar{A}(\gamma_2)\cdot 
\bar{A}(\gamma_1)$, where $\gamma^{-1}$ is obtained from $\gamma$ 
by reversing its orientation, $\gamma_2\circ \gamma_1$ denotes 
the composition of the two curves (obtained by connecting the end 
of $\gamma_1$ with the beginning of $\gamma_2$) and 
$\bar{A}(\gamma_2)\cdot\bar{A}(\gamma_1)$ is the composition in 
$SU(2)$.  The space of such generalized connections is denoted 
$\overline{\cal A}$. The cylindrical functions $\Psi_{\Gamma, f}(A)$ 
defined in section \ref{states} as functions on the space of smooth 
connections extend immediately to generalized connections
\begin{equation}
	\Psi_{\Gamma,f}(\bar A)=f(\bar A(\gamma_{1},\ldots,\bar 
	A(\gamma_{n})).
\end{equation}

We can define a measure $d\mu_{0}$ on the space of generalized 
connections $\overline{\cal A}$ by
\begin{equation}
	\int d\mu_{0}[\bar A]\, \Psi_{\Gamma,f}(\bar A)
	\equiv \int_{SU(2)^{n}} \!\!\!\! 
	dg_{1}\ldots dg_{n\,} f(g_{1}\ldots g_{n}).
	\label{measure}
\end{equation}

In fact, one may show that (\ref{measure}) defines (by linearity 
and continuity) a well-defined absolutely continuous measure on 
$\overline{\cal A}$.  This is the Ashtekar-Lewandowski (or 
Ashtekar-Lewandowski-Baez) measure 
\cite{AshtekarLewandowski,AshtekarLewandowski3,%
AshtekarLewandowski2,Baez}.  Then, one can prove that ${\cal H} = 
L_{2}[\overline{\cal A},d\mu_0]$, under the natural isomorphism 
given by identifying cylindrical functions.  It follows 
immediately that the transformation (\ref{looptransform1}) between 
the connection representation and the ``old'' loop representation 
is given by
\begin{equation}
	\psi(\alpha)=\int d\mu_{0}[\bar A] \ 
	\rm{Tr} \overline{{\cal P} e^{\int_{\alpha}A}}	\Psi(\bar A).
	\label{looptransform}
\end{equation}
This is the loop transform formula that was derived heuristically 
in \cite{RovelliSmolin90}; here it becomes rigorously defined. 

Furthermore, $\cal H$ can be seen as the projective limit of the 
projective family of the Hilbert spaces ${\cal H}_{\Gamma}$, 
associated to each graph $\Gamma$ immersed in $M$.  ${\cal 
H}_{\Gamma}$ is defined as the space $L_{2}[SU(2)^{n}, dg_{1} 
\ldots dg_{n}]$, where $n$ is the number of links in $\Gamma$. 
The cylindrical function $\Psi_{\Gamma,f}(A)$ is naturally 
associated to the function $f$ in ${\cal H}_{\Gamma}$, and the
projective structure is given by the natural map (\ref{prmap})  
\cite{AshtekarEtAl95,MarolfMourao}. 

Finally, Ashtekar and Isham \cite{AshtekarIsham} have recovered 
the representation of the loop algebra by using C*-algebra 
representation theory: The space $\overline{\cal A}/{\cal G}$, 
where $\cal G$ is the group of local $SU(2)$ transformations 
(which acts in the obvious way on generalized connections), is 
precisely the Gelfand spectrum of the abelian part of the loop 
algebra.  One can show that this is a suitable norm closure of 
the space of smooth $SU(2)$ connections over physical space, 
modulo gauge transformations.

Thus, a number of powerful mathematical tools are at hand for 
dealing with nonperturbative quantum gravity.   These 
include: Penrose's spin network theory, $SU(2)$ 
representation theory, Kauffman tangle theoretical recoupling 
theory, Temperley-Lieb algebras, Gelfand's $C^*$algebra 
spectral representation theory, infinite dimensional measure 
theory and differential geometry over infinite dimensional 
spaces.

\subsection{Diffeomorphism invariance}\label{diffinvariance}

The next step in the construction of the theory is to factor 
away diffeomorphism invariance.  This is a key step for two 
reasons.  First of all, $\cal H$ is a ``huge'' non-separable 
space.  It is far ``too large'' for a quantum field theory.  
However, most of this redundancy is gauge, and disappears 
when one solves the diffeomorphism constraint, defining the 
diff-invariant Hilbert space ${\cal H}_{Diff}$.  This is the reason 
for which the loop representation, as defined here, is of 
great value in diffeomorphism invariant theories only.

The second reason is that ${\cal H}_{Diff}$ turns out to have a 
natural basis labeled by knots.  More precisely by 
``s-knots''.  An s-knot $s$ is an equivalence class of spin 
networks $S$ under diffeomorphisms.  An s-knot is 
characterized by its ``abstract'' graph (defined only by the 
adjacency relations between links and nodes), by the 
coloring, and by its knotting and linking properties, as in 
knot-theory. Thus, the physical quantum states of the 
gravitational field turn out to be essentially classified by 
knot theory.

There are various equivalent way of obtaining ${\cal H}_{Diff}$ 
from ${\cal H}$.  One can use regularization techniques for 
defining the quantum operator corresponding to the classical 
diffeomorphism constraint in terms of elementary loop 
operators, and then find the kernel of such operator.  
Equivalently, one can factor ${\cal H}$ by the natural action 
of the Diffeomorphism group that it carries.  Namely 
\begin{equation} 
{\cal H}_{Diff}={{\cal H}\over Diff(M)}.  
\end{equation} 
There are several rigorous ways for defining the quotient of a 
Hilbert space by the unitary action of a group.  See in 
particular the construction in \cite{AshtekarEtAl95}, which 
follows the ideas of Marolf and Higuchi 
\cite{MarolfGroup,MarolfGroup2,MarolfGroup3,Higuchi}.

In the quantum gravity literature, a big deal has been made of 
the problem that a scalar product is not defined on the space of 
solutions of a constraint $\hat C$, defined on a Hilbert space $\cal 
H$.  This, however, is a false problem.  It is true that if zero 
is in the continuum spectrum of $\hat C$ then the corresponding 
eigenstates are generalized states and the $\cal H$ scalar 
product is not defined between them.  But the generalized 
eigenspaces of $\hat C$, including the kernel, inherit {\em 
nevertheless\/} a scalar product from $\hat H$.  This can be seen 
in a variety of equivalent ways.  For instance, it can be seen  
from the following theorem.  If $\hat C$ is self adjoint, then 
there exist a measure $\mu(\lambda)$ on its spectrum and a family 
of {\em Hilbert\/} spaces ${\cal H}(\lambda)$ such that
\begin{equation}
	{\cal H} = \int d\mu(\lambda) {\cal H}(\lambda),
\end{equation}
where the integral is the continuous sum of Hilbert spaces 
described for instance in \cite{Guichardet}. Clearly ${\cal 
H}(0)$ is the kernel of $\hat C$ {\em equipped with a scalar 
product.} This is discussed for instance in \cite{RovelliTh}. 

There are two distinct ways of factoring away the diffeomorphisms 
in the quantum theory, yielding two distinct version of the 
theory.  The first way is to factor away smooth transformations of 
the manifold.  In doing so, finite dimensional moduli spaces 
associated with high valence nodes appear 
\cite{GrottRovelli96}, so that the resulting Hilbert space is still 
nonseparable.  The physical relevance of these moduli parameters 
is unclear at this stage, since they do not seem to play any role 
in the quantum theory.  Alternatively, one can consistently 
factor away {\em continuous\/} transformations of the manifold. 
This possibility has been explored by Zapata in \cite{Zapata,Zapata2}, 
and seems to lead to a consistent theory free of the residual non 
separability. 

\subsection{Dynamics}

Finally, the definition of the theory is completed by giving the 
hamiltonian constraint.  A number of approaches to the definition 
of a hamiltonian constraint have been attempted in the past, with 
various degrees of success.  Recently, however, Thiemann has 
succeeded in providing a regularization of the hamiltonian 
constraint that yields a well defined, finite operator.  
Thiemann's construction \cite{Thiemann96,Thiemann96b,Thiemann96c} 
is based on several clever ideas.  I will not describe it here.  
Rather, I will sketch below the final form of the constraint (for 
the Lapse=1 case), following \cite{Rovelli95b}.

I begin with the Euclidean hamiltonian constraint. We have 
 \begin{equation}
	\hat H |s \rangle = \sum_i \sum_{(IJ)}\, \sum_{\epsilon=\pm 
	1} \, \sum_{\epsilon'=\pm 1}\, 
	A_{\epsilon\epsilon'}(p_i...p_n) \, \hat 
	D_{i;(IJ),\epsilon\epsilon'}\, |s\rangle.
\label{lee}
\end{equation}
Here $i$ labels the nodes of the s-knot $s$; $(IJ)$ labels 
couples of (distinct) links emerging from $i$.  $p_1...p_n$ are 
the colors of the links emerging from $i$.  $\hat 
D_{i;(IJ)\epsilon\epsilon'}$ is the operator that acts on an 
$s\,$-knot by: (i) creating two additional nodes, one along 
each of the two links $I$ and $J$; (ii) creating a novel link, 
colored 1, joining these two nodes, (iii) assigning the coloring 
$p_I+\epsilon$ and, respectively, $p_J+\epsilon'$ to the links 
that join the new formed nodes with the node $i$.  This is 
illustrated in Figure 2.  \vskip.5cm
\begin{figure}
\centerline{
\setlength{\unitlength}{.7pt}
\begin{picture}(344,114)(105,606)
\thicklines
\put(201,648){\line(-5,-3){ 60}}
\put(201,648){\line( 5,-3){ 60}}
\put(201,648){\line( 0, 1){ 60}}
\put(162,618){\makebox(0,0)[lb]{\raisebox{0pt}[0pt][0pt]{
 r}}}
\put(185,694){\makebox(0,0)[lb]{\raisebox{0pt}[0pt][0pt]{
 q}}}
\put(234,618){\makebox(0,0)[lb]{\raisebox{0pt}[0pt][0pt]{
 p}}}
\put(95,654){\makebox(0,0)[lb]{\raisebox{0pt}[0pt][0pt]{
\^D${}_{+-}$}}}
\put(371,685){\line( 2,-3){ 40.462}}
\put(371,648){\line(-5,-3){ 60}}
\put(371,648){\line( 5,-3){ 60}}
\put(371,648){\line( 0, 1){ 60}}
\put(271,654){\makebox(0,0)[lb]{\raisebox{0pt}[0pt][0pt]{
=}}}
\put(326,609){\makebox(0,0)[lb]{\raisebox{0pt}[0pt][0pt]{
 r}}}
\put(422,606){\makebox(0,0)[lb]{\raisebox{0pt}[0pt][0pt]{
 p}}}
\put(359,703){\makebox(0,0)[lb]{\raisebox{0pt}[0pt][0pt]{
 q}}}
\put(403,657){\makebox(0,0)[lb]{\raisebox{0pt}[0pt][0pt]{
 1}}}
\put(343,657){\makebox(0,0)[lb]{\raisebox{0pt}[0pt][0pt]{
q$-$1}}}
\put(362,621){\makebox(0,0)[lb]{\raisebox{0pt}[0pt][0pt]{
p$+$1}}}
\end{picture}
}
\caption{Action of $\hat D_{i;(IJ)\epsilon\epsilon'}$.}
\end{figure}
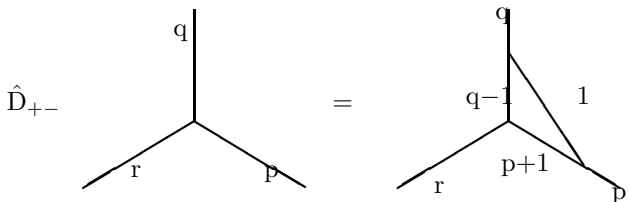

The coefficients $A_{\epsilon\epsilon'}(p_i...p_n)$, which are 
finite, can be expressed explicitly (but in a rather laborious 
way) in terms of products of linear combinations of $6-j$ symbols 
of $SU(2)$, following the techniques developed in detail in 
\cite{DePietriRovelli}.  Some of these coefficients have been 
explicitly computed \cite{BorissovEtAl97}.  The Lorentzian 
hamiltonian constraint is given by a similar expression, but 
quadratic in the $\hat D$ operators.

The operator defined above is obtained by introducing a 
regularized expression for the classical hamiltonian constraint, 
written in terms of elementary loop observables, turning these 
observables into the corresponding operators and taking the 
limit.  The construction works rather magically, relying on the 
fact, first noticed in \cite{RovelliSmolin94b}, that certain 
operator limits $\hat O_{\epsilon}\rightarrow \hat O$ turns out 
to be finite on diff invariant states, thanks to the fact that 
for $\epsilon$ and $\epsilon'$ sufficiently small, $\hat 
O_{\epsilon}|\Psi\rangle$ and $\hat O_{\epsilon'}|\Psi\rangle$ 
are diffeomorphic equivalent.  Thus, here diff invariance plays 
again a crucial role in the theory.

For a discussion of the problems raised by the Thiemann operator 
and of the variant proposed, see section \ref{7}.

\subsection{Unfreezing the frozen time formalism: the 
covariant form of loop quantum gravity}\label{spacetime}

A recent development in the formalism is the translation 
of loop quantum gravity into spacetime covariant form.  This 
was initiated in \cite{RovelliSurf,ReisenbergerRovelli} by following 
the steps that Feynman took in defining path integral 
quantum mechanics starting from the Schr\"odinger canonical 
theory.  More precisely, it was proven in 
\cite{ReisenbergerRovelli} that the matrix elements of the 
operator $U(T)$ 
\begin{equation}
	U(T) \equiv e^{\int_{0}^{T}dt\ \int d^{3}x\ \hat H(x)}, 
\end{equation}
obtained exponentiating the (Euclidean) hamiltonian 
constraint in the proper time gauge (the operator that 
generates evolution in proper time) can be expanded in a 
Feynman sum over paths.  In conventional QFT each term of 
a Feynman sum corresponds naturally to a certain Feynman 
diagram, namely a set of lines in spacetime meeting at 
vertices (branching points).  A similar 
natural structure of the terms appears in quantum gravity, 
but surprisingly the diagrams are now given by {\em surfaces\/} is 
spacetime that branch at vertices.  Thus, one has a 
formulation of quantum gravity as a sum over surfaces in 
spacetime. Reisenberger \cite{Reisenberger94} and Baez 
\cite{Baez94b} had argued in the past that such a 
formulation should exist, and Iwasaki has developed a similar 
construction in 2+1 dimensions. Intuitively, the time evolution of a 
spin network in spacetime is given by a colored surface. 
 The surfaces capture the gravitational degrees 
of freedom.  The formulation is ``topological'' in the 
sense that one must sum over topologically inequivalent 
surfaces only, and the contribution of each surface 
depends on its topology only.  This contribution is given 
by the product of elementary ``vertices'', namely points 
where the surface branches. 

The transition amplitude between two s-knot states 
$|s_{i}\rangle$ and $|s_{f}\rangle$ in a proper time $T$ is given 
by summing over all (branching, colored) surfaces $\sigma$ 
that are bounded by the two s-knots $s_{i}$ and $s_{f}$ 
 \begin{equation}
\langle s_{f}|U(T)|s_{i}\rangle  =
\sum_{\stackrel{\scriptstyle\sigma}
{\scriptstyle\partial\sigma=s_i\cup s_f}}
 {\cal A}[\sigma](T)
\label{uno}. 
 \end{equation}
The weight ${\cal A}[\sigma](T)$ of the surface $\sigma$ is given 
by a product over the $n$ vertices $v$ of $\sigma$: 
\begin{equation} 
{\cal A}[\sigma](T)= {(\mbox{\rm - 
}\!i\;T)^{n}\over n!}\ \prod_{v} A_v(\sigma).
\label{due}
\end{equation} 
The contribution $A_v(\sigma)$ of each vertex is given by the 
matrix elements of the hamiltonian constraint operator between the 
two s-knots obtained by slicing $\sigma$ immediately below and 
immediately above the vertex.  They turn out to depend only on the 
colors of the surface components immediately adjacent the vertex 
$v$.  The sum turns out to be finite and explicitly computable 
order by order.

As in the usual Feynman diagrams, the vertices describe the 
elementary interactions of the theory.  In particular, here one 
sees that the complicated structure of the Thiemann hamiltonian, 
which makes a node split into three nodes, corresponds to a 
geometrically very simple vertex.  Figure 3 is a picture of the 
elementary vertex.  Notice that it represents nothing but the 
spacetime evolution of the elementary action of the hamiltonian 
constraint, given in Figure 2.

\begin{figure} 
\centerline{\mbox{\epsfig{file=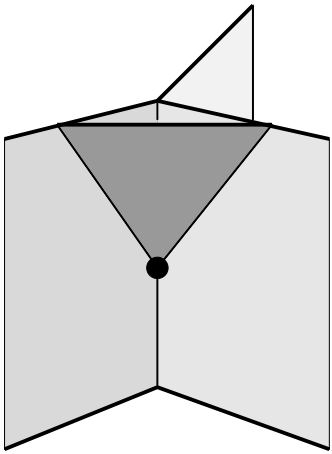}}} 
\caption{The elementary vertex.} 
\end{figure}

An example of a surface in the sum is given in Figure 4.

\begin{figure} \centerline{\mbox{\epsfig{file=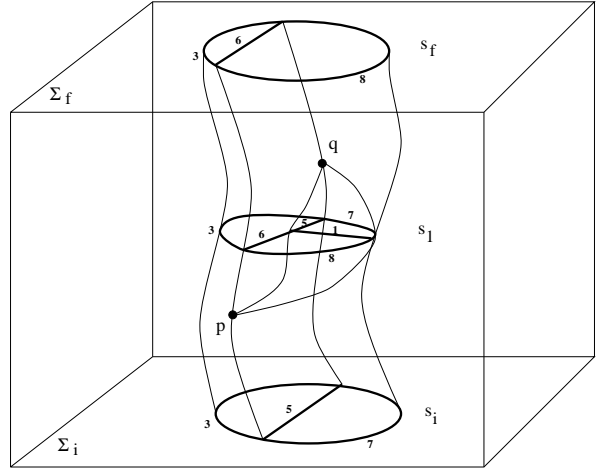}}}
 \caption{A term of second order.}
\end{figure}

The sum over surfaces version of loop quantum gravity provides a 
link with certain topological quantum field theories and in 
particular with the the Crane-Yetter model 
\cite{Crane1,Crane2,CraneFrenkel,CraneKauffmanYetter,CraneYetter}, 
which admit an extremely similar representation.  For a 
discussion on the precise relation between topological quantum 
field theory and diffeomorphism invariant quantum field theory, 
see \cite{ReisenbergerRovelli} and 
\cite{RovelliPonzano,Iwasaki,Foxon}.

The idea of expressing the theory as a sum over surfaces has been 
developed by Baez \cite{Baez97}, who has studied the general form 
of generally covariant quantum field theories formulated in this 
manner, and by Smolin and Markopoulou \cite{SmolinMarkopoulo}, 
who have studied how to directly capture the Lorentzian causal 
structure of general relativity modifying the elementary 
vertices.  They have also explored the idea that the long range 
correlations of the low energy regime of the theory are related 
to the existence of a phase transition in the microscopic 
dynamics, and has found intriguing connections with the 
theoretical description of percolation.

\section{Physical results}\label{6}

In section \ref{5}, I have sketched the basic structure of loop 
quantum gravity.  This structure has been developed in a number 
of directions, and has been used to derive a number of results.  
Without any ambition of completeness, I list below some of these 
developments.

\subsection{Technical} \label{technical}

\begin{itemize}

\item {\bf Solutions of the hamiltonian constraints.} One of 
the most surprising results of the theory is that it has been 
possible to find exact solutions of the hamiltonian constraint.  
This follows from the key result that the action of the 
hamiltonian constraints is non vanishing only over nodes of 
the s-knots \cite{RovelliSmolin88,RovelliSmolin90}.  Therefore 
s-knots without nodes are physical states that solve the 
quantum Einstein dynamics.  There is an infinite number of 
independent states of this sort, classified by conventional knot 
theory.  The physical interpretation of these solutions is still 
rather obscure.   Various other solutions have been found.  See 
the recent review \cite{Ezawa} and reference therein.  See also 
\cite{Husain,Kodama,Bruegmann2,H5,H4,H6,H7,GambiniPullin,Kauffman94b}. 
In particular, Pullin has 
studied in detail solutions related to the Chern-Simon term in 
the connection representation and to the Jones polynomial in the 
loop representation. According to a celebrated result by Witten 
\cite{Witten}, the two are the loop transform of each other. 

\item {\bf Time evolution.  Strong field perturbation expansion.  
``Topological Feynman rules''.  } Trying to describe the temporal 
evolution of the quantum gravitational field by solving the 
hamiltonian constraint yields the conceptually well-defined 
\cite{RovelliTime}, but notoriously non-transparent 
frozen-time formalism.  An alternative is to study the evolution 
of the gravitational degrees of freedom with respect to some 
matter variable, coupled to the theory, which plays the role of a 
phenomenological ``clock''.  This approach has lead to the 
tentative definition of a physical hamiltonian 
\cite{RovelliSmolin94b,H9}, 
and to a preliminary investigation of the possibility of 
transition amplitudes between s-knot states, order by order in a 
(strong coupling) perturbative expansion \cite{Rovelli95b}.  In 
this context, diffeomorphism invariance, combined with the key 
result that the hamiltonian constraint acts on nodes only, 
implies that the ``Feynman rules'' of such an expansion are purely 
topological and combinatorial.

\item {\bf Fermions.  } Fermions have been added to the theory 
\cite{MoralesRovelli,MoralesRovelli2,BaezKrasnov,ThiemannFermions}.  
Remarkably, all the 
important results of the pure GR case survive in the GR+fermions 
theory.  Not surprisingly, fermions can be described as open ends 
of ``open spin networks''.

\item {\bf Maxwell and Yang-Mills.  } The extension of the theory to 
the Maxwell field has been studied in 
\cite{Krasnov95,GambiniPullin93}.  The extension to Yang-Mills theory 
has been recently explored in \cite{ThiemannYM}.  In 
\cite{ThiemannYM}, Thiemann shows that the Yang-Mills term in the 
quantum hamiltonian constraint can be defined in a rigorous manner, 
extending the methods of \cite{Thiemann96,Thiemann96b,Thiemann96c}.  A 
remarkable result in this context is that ultraviolet divergences do 
not seem to appear, strongly supporting the expectation that the 
natural cut off introduced by quantum gravity might cure the 
ultraviolet difficulties of conventional quantum field theory.

\item {\bf Application to other theories.  } The loop 
representation has been applied in various other contexts such as 
2+1 gravity \cite{AshtekarEtAl89,Marolf93,AshtekarLoll} (on 2+1 
quantum gravity, in the loop and in other representations, see 
\cite{Carlip}) and others \cite{AshtekarRovelli}.

\item {\bf Lattice and simplicial models.  } A number of 
interesting discretized versions of the theory are being studied.  
See in particular 
\cite{Loll95a,Reisenberger,GambiniPullin,GambiniLattice}.

\end{itemize}
 
       \subsection{Physical}\label{physical}

\begin{itemize}

\item {\bf Planck scale discreteness of space}

The most remarkable physical result obtained from loop quantum gravity 
is, in my opinion, evidence for a physical (quantum) discreteness of 
space at the Planck scale.  This is manifested in the fact that 
certain operators corresponding to the measurement of geometrical 
quantities, in particular area and volume, have discrete spectrum.  
According to the standard interpretation of quantum mechanics (which 
we adopt), this means that the theory predicts that a physical 
measurement of an area or a volume will necessarily yield quantized 
results.  Since the smallest eigenvalues are of Planck scale, this 
implies that there is no way of observing areas or volumes smaller 
than Planck scale.  Space comes in ``quanta'' in the same manner as 
the energy of an oscillator.  The spectra of the area and volume 
operators have been computed with much detail in loop quantum gravity.  
These spectra have a complicated structure, and they constitute 
detailed quantitative physical predictions of loop quantum gravity on 
Planck scale physics.  If we had experimental access to Planck scale 
physics, they would allow the theory to be empirically tested in great 
detail.

A few comments are in order.
\begin{itemize}

	\item The result of the discreteness of area and volume is due 
	to Rovelli and Smolin, and appeared first in reference 
	\cite{RovelliSmolin95}.  Later, the result has been recovered 
	by alternative techniques and extended by a number of authors.  
	In particular, Ashtekar and Lewandowski 
	\cite{AshtekarLewandowskiArea2} have repeated the derivation, 
	using the connection representation, and have completed the 
	computation of the spectrum (adding the sector which was not 
	computed in \cite{RovelliSmolin95}.)  The Ashtekar-Lewandowski 
	component of the spectrum has then been rederived in the loop 
	representation by Frittelli Lehner and Rovelli in 
	\cite{FrittelliEtAl}.  Loll has employed lattice techniques to 
	point out a numerical error in \cite{RovelliSmolin95} 
	(corrected in the Erratum) in the eigenvalues of the volume.  
	The analysis of the volume eigenvalues has been performed in 
	\cite{DePietriRovelli}, where general techniques for 
	performing these calculations are described in detail.  The 
	spectrum of the volume has then been analyzed also in 
	\cite{ThiemannVolume}.  There are also a few papers that have 
	anticipated the main result presented in 
	\cite{RovelliSmolin95}.  In particular, Ashtekar Rovelli and 
	Smolin have argued for a physical discreteness of space 
	emerging from the loop representation in \cite{weave}, where 
	some of the eigenvalues of the area already appear, although 
	in implicit form.  The first explicit claim that area 
	eigenvalues might in principle be observable (in the presence 
	of matter) is by Rovelli in \cite{RovelliArea}.

\item The reader will wonder why area and volume seem to play here a 
role more central than length, when classical geometry is usually 
described in terms of lengths.  The reason is that the length operator 
is difficult to define and of more difficult physical interpretations.  
For attempts in this direction, see \cite{ThiemannLength}.  Whether 
this is simply a technical difficulty or it reflects some deep fact, 
is not clear to me.\footnote{I include here a comment on this issue 
I received from John Baez: ``I believe there {\em is\/} a deep reason
why area is more fundamental than length in loop quantum gravity.  One
way to say it is that the basic field is not the tetrad $e$ but the
2-form $E = e\wedge e$.  Another way to say it is that the loop
representation is based on the theory of quantized angular momentum.
Angular momentum is not a vector but a bivector, so it corresponds not
to an arrow but to a oriented area element.''  On this, see 
Baez's \cite{Baez97}.  On the relation between $E$ field and area see in 
particular \cite{Rovelli93b}.}

	\item Area and volume are not gauge invariant operators.  
	Therefore, we cannot directly interpret them as representing 
	physical measurements.  Realistic physical measurements of 
	areas and volumes always refer to {\em physical\/} surfaces 
	and spatial regions, namely surfaces and spatial regions 
	determined by some physical object.  For instance, I can 
	measure the area of the surface of a certain table.  In the 
	dynamical theory that describes the gravitational field as 
	well as the table, the area of the surface of the table is a 
	diffeomorphism invariant quantity $A$, which depends on 
	gravitational as well as matter variables.  In the quantum 
	theory, $A$ will be represented by a diffeomorphism invariant 
	operator.  Now, as first realized in \cite{RovelliArea}, it 
	is plausible that that the operator $A$ is, mathematically, 
	the same operator as the pure gravity area operator.  This is 
	because we can gauge fix the matter variables, and use matter 
	location as coordinates, so that non-diff-invariant 
	observables in the pure gravity theory correspond precisely 
	to diff-invariant observables in the matter+gravity theory.  
	Thus, discreteness of the spectrum of the area operator is 
	likely to imply discreteness of physically measurable areas, 
	but it is important to emphasize that this implication is 
	based on some additional hypothesis on the relation between 
	the pure gravity and the gravity+matter theories.

\end{itemize}

 The discreteness of area and volume is derived as follows.  Consider 
 the area $A$ of a surface $\Sigma$.  The physical area $A$ of 
 $\Sigma$ depends on the metric, namely on the gravitational field.  
 In a quantum theory of gravity, the gravitational field is a quantum 
 field operator, and therefore we must describe the area of $\Sigma$ 
 in terms of a quantum observable, described by an operator $\hat A$.  
 We now ask what is the quantum operator $\hat A$ in nonperturbative 
 quantum gravity.  The result can easily be worked out by writing the 
 standard formula for the area of a surface, and replacing the metric 
 with the appropriate function of the loop variables.  Promoting these 
 loop variables to operators, we obtain the area operator $\hat A$.  
 The actual construction of this operator requires regularizing the 
 classical expression and then taking the limit of a sequence of 
 operators, in a suitable operator topology.  For the details of this 
 construction, see 
 \cite{RovelliSmolin95,DePietriRovelli,FrittelliEtAl,BorissovEtAl97}.  
 An alternative regularization technique is discussed in 
 \cite{AshtekarLewandowskiArea2}.  The resulting area operator $\hat 
 A$ acts as follows on a spin network state $|S\rangle$ (assuming here 
 for simplicity that $S$ is a spin network without nodes on $\Sigma$):
\begin{equation}
\hat{A}[\Sigma]\ |S\rangle=
\left({l^2_0\over 2} \sum_{i\in\{S\cap\Sigma\}}
\sqrt{p_i(p_i+2)}\right) \  |S\rangle
\end{equation}
where $i$ labels the intersections between the spin network $S$ 
and the surface $\Sigma$, and $p_{i}$ is the color of the link of 
$S$ crossing the $i-th$ intersection.

This result shows that the spin network states (with
a finite number of intersection points with the surface
and no nodes on the surface) are eigenstates of the
area operator. The corresponding spectrum is labeled by
multiplets  $\vec p = (p_1, ..., p_n)$ of positive half
integers, with arbitrary $n$, and given by
\begin{equation}
   A_{\vec p}\,[\Sigma] = {l^2_0\over 2} \, \sum_i  
   \sqrt{p_i(p_i+2)}.
\end{equation}
Shifting from color to spin notation reveals the $SU(2)$ origin
of the spectrum:
\begin{equation}
   A_{\vec j}\,[\Sigma] = l^2_0 \, \sum_i  \sqrt{j_i(j_i+1)},
\label{spec}
\end{equation}
where $j_1, ..., j_n$ are half integer.  For the full spectrum, 
see \cite{AshtekarLewandowskiArea2} (connection representation) 
and \cite{FrittelliEtAl} (loop representation).

A similar result can be obtained for the volume 
\cite{RovelliSmolin95,Loll95b,Loll95bb,%
DePietriRovelli,Lewandowski97}. Let us restrict here for
simplicity to spin networks $S$ with nondegenerate fourvalent 
nodes, labeled by an index $i$.  Let 
$a_i,b_i,c_i,d_i$ be the colors of the links adjacent to the
$i-th$ node and let $J_i$ label the basis in the intertwiner
space.  The volume operator $\hat V$ acts as follows
\begin{equation}
\hat V |S\rangle  = \frac{l_0^3}{2} \sum_{i}
  \sqrt{ | i W^{(i)}(a_i,b_i,c_i,d_i) |} |S\rangle
\end{equation}
where $W^{i}$ is an operator that acts of the finite dimensional 
space of the intertwiners in the $i-th$ node, and its matrix 
elements are explicitly given (in a suitable basis) by 
($\epsilon=\pm$)
\begin{eqnarray} 
&& \!\!\! W^{i}(a,b,c,d){}_{t-\epsilon}^{t+\epsilon} 
=  - \epsilon (-1)^{\frac{a+b+c+d}{2}}\   \times
\nonumber \\ &&      
\     \bigg[ \frac{1}{4 t (t+2)}
      \frac{a+b+t+3}{2}\frac{c+d+t+3}{2}
\nonumber \\ &&  
\ \  \    \frac{1+a+b-t}{2}\frac{1+a+t-b}{2}\frac{1+b+t-a}{2}
\label{v0}
\\ &&  
\ \ \      \frac{1+c+d-t}{2}\frac{1+c+t-d}{2}\frac{1+d+t-c}{2}
     \bigg]^{\frac{1}{2}}.
\nonumber
\end{eqnarray}
See \cite{BorissovEtAl97}.  The volume eigenvalues $v_{i}$ are 
obtained by diagonalizing these matrices.  For instance, in the 
simple case $a=b$, $c=d=1$, we have
\begin{equation}
	v_{a}=\frac{(\hbar G)^{3}}{2}\sqrt{a(a+2)};
	\label{v1}
\end{equation}
if $d=a+b+c$, we have
\begin{equation}
	v_{a,b,c}=\frac{(\hbar G)^{3}}{2}\sqrt{abc(a+b+c)}. 
\label{v2}
\end{equation}
For more details, and the full derivation of these formulas, see 
\cite{BorissovEtAl97,ThiemannVolume}

\item {\bf Classical limit.  Quantum states representing flat 
spacetime.  Weaves.  Discrete small scale structure of space.} 

The s-knot states do not represent excitations of the quantum 
gravitational field over flat space, but rather over 
``no-space'', or over the $g_{\mu\nu}=0$ solution.  A natural 
problem is then how flat space (or any other smooth geometry) 
might emerge from the theory.  Notice that in a general 
relativistic context the Minkowski solution does not have all the 
properties of the conventional field theoretical vacuum.  (In 
gravitational physics there is no real equivalent of the 
conventional vacuum, particularly in the spatially compact case.)  
One then expects that flat space is represented by some highly 
excited state in the theory.  States in $\cal H$ that describe 
flat space when probed at low energy (large distance) have been 
studied in \cite{weave,Zegwaard,Borissov,GrotRovelli}.  These 
have a discrete structure at the Planck scale.  Furthermore, 
small excitations around such states have been considered in 
\cite{IwasakiRovelli}, where it is shown that $\cal H$ contains 
all ``free graviton'' physics, in a suitable approximation.

\item {\bf The Bekenstein-Mukhanov effect.}

Recently, Bekenstein and Mukhanov \cite{BekensteinMukhanov} have 
suggested that the thermal nature of Hawking's radiation 
\cite{Hawking,Hawking2} may be affected by quantum properties of 
gravity (For a review of earlier suggestions in this direction, 
see \cite{SmolinMoG}).  Bekenstein and Mukhanov observe that in 
most approaches to quantum gravity the area can take only 
quantized values \cite{Garay}.  Since the area of the black hole 
surface is connected to the black hole mass, black hole mass is 
likely to be quantized as well.  The mass of the black hole 
decreases when radiation is emitted.  Therefore emission happens 
when the black hole makes a quantum leap from one quantized value 
of the mass (energy) to a lower quantized value, very much as 
atoms do.  A consequence of this picture is that radiation is 
emitted at quantized frequencies, corresponding to the 
differences between energy levels.  Thus, quantum gravity implies 
a discretized emission spectrum for the black hole radiation.

This result is not physically in contradiction with Hawking's 
prediction of a continuous thermal spectrum, because spectral 
lines can be very dense in macroscopic regimes.  But Bekenstein 
and Mukhanov observed that if we pick the simplest ansatz for the 
quantization of the area --that the Area is quantized in multiple 
integers of an elementary area $A_0$--, then the emitted spectrum 
turns out to be macroscopically discrete, and therefore very 
different from Hawking's prediction.  I will denote this effect as 
the kinematical Bekenstein-Mukhanov effect.  Unfortunately, 
however, the kinematical Bekenstein-Mukhanov effect disappears if 
we replace the naive ansatz with the spectrum (\ref{spec}) 
computed from loop quantum gravity.  In loop quantum gravity the 
eigenvalues of the area become exponentially dense for a 
macroscopic black hole, and therefore the emission spectrum can be 
consistent with Hawking's thermal spectrum.  This is due to the 
details of the spectrum (\ref{spec}) of the area.  A detailed 
discussion of this result is in \cite{BarreiraEtAl}, but the 
result was already contained (implicitly, in the first version) 
in \cite{AshtekarLewandowskiArea2}.  It is important to notice that 
the density of the eigenvalues shows only that the simple 
kinematical argument of Bekenstein and Mukhanov is not valid in 
this theory, and not that their conclusions is necessarily wrong.  
As emphasized by Mukhanov, a discretization of the emitted 
spectrum could be still be originated dynamically.

\item {\bf Black Hole Entropy from Loop Quantum Gravity}

Indirect arguments \cite{Hawking,Hawking2,Bekenstein,Wald} 
strongly support the idea that a Schwarzschild black hole of 
(macroscopic) area $A$ behaves as a thermodynamical system 
governed by the Bekenstein-Hawking entropy
  \begin{equation}
                 S = {k\over 4 \hbar G_{Newton}} \ A
   \label{bh}
  \end{equation}
($k$ is the Boltzmann constant; here I put the speed of light 
equal to one, but write the Planck and Newton constants 
explicitly).  A physical understanding and a first principles  
derivation of this relation requires quantum gravity, and 
therefore represents a challenge for every candidate theory of 
quantum theory.  A derivation of the Bekenstein-Hawking 
expression (\ref{bh}) for the entropy of a Schwarzschild black 
hole of surface area $A$ via a statistical mechanical 
computation, using loop quantum gravity, was obtained in 
\cite{Krasnov,Krasnov2,Rovelli96}.

This derivation is based on the ideas that the entropy of the 
hole originates from the microstates of the horizon that 
correspond to a given macroscopic configuration 
\cite{York,Carlip95,Carlip97,Bala2,Bala}.  Physical arguments 
indicate that the entropy of such a system is determined by an 
ensemble of configurations of the horizon with fixed area 
\cite{Rovelli96}.  In the quantum theory these states are finite 
in number, and can be counted \cite{Krasnov,Krasnov2}.  Counting 
these microstates using loop quantum gravity yields
\begin{equation}
    S = \frac{c}{\gamma}\ {k\over 4 \hbar G_{Newton}} \ A.
\end{equation}
(An alternative derivation of this result has been announced from 
Ashtekar, Baez, Corichi and Krasnov \cite{AshtekarEtAl97}.)\ 
$\gamma$ is defined in section \ref{5}, and $c$ is a real number 
of the order of unity that emerges from the combinatorial 
calculation (roughly, $c \sim 1/4\pi$).  If we choose $\gamma=c$, 
we get (\ref{bh}) \cite{RovelliThiemann,CorichiKrasnov}.  Thus, 
the theory is compatible with the numerical constant in the 
Bekenstein-Hawking formula, but does not lead to it univocally.  
The precise significance of this fact is under discussion.  In 
particular, the meaning of $\gamma$ is unclear.  Jacobson has 
suggested \cite{Jacobson} that finite renormalization effects may 
affect the relation between the bare and the effective Newton 
constant, and this may be reflected in $\gamma$.  For discussion 
of the role of $\gamma$ in the theory, see 
\cite{RovelliThiemann}.  On the issue of entropy in loop gravity, 
see also \cite{Smolin95}.

\end{itemize}
  
\section{Main open problems and main current lines of 
investigation}\label{7}

\begin{description}

\item[Hamiltonian constraint.] 

The kinematic of the theory is well understood, both physically 
(quanta of area and volume, polymer-like geometry) and from the 
mathematical point of view ($\cal H$, s-knot states, area and 
volume operators).  The part of the theory which is not yet fully 
under control is the dynamics, which is determined by the 
hamiltonian constraint.  A plausible candidate for the quantum 
hamiltonian constraint is the operator introduced by Thiemann 
\cite{Thiemann96,Thiemann96b,Thiemann96c}.  The commutators of 
the Thiemann operator with itself and with the diffeomorphism 
constraints close, and therefore the operator defines a complete 
and consistent quantum theory.  However, doubts have been raised 
on the physical correctness of this theory, and some variants of 
the operator have been considered.

The doubts originate from various considerations.  First, Lewandowski, 
Marolf and others have stressed the fact the quantum constraint 
algebra closes, but it is not isomorphic to the classical constraint 
algebra of GR \cite{LewandowskiMarolf}.  Recently, a detailed analysis 
of this problem has been completed by Marolf, Lewandowski, Gambini and 
Pullin \cite{LewandowskiEtAl}.  The failure to reproduce the classical 
constraint algebra has been disputed, and is not necessarily a 
problem, since the only strict requirement on the quantum theory, 
besides consistency, is that its {\em gauge invariant\/} physical 
predictions match the ones of classical general relativity in the 
appropriate limit.  Still, the difference in the algebras may be seen as 
circumstantial evidence (not a proof) for the failure of the classical 
limit.  The issue is technically delicate and still 
controversial.  I hope I will be able to say something more 
definitive in the next update of this review. 

Second, Br\"ugmann \cite{Bruegmann3} and Smolin \cite{Smolin96} 
have pointed out a sort of excess ``locality'' in the form of the 
operator, which, intuitively, seems in contradiction with the 
propagation properties of the Einstein equations.  Finally, by 
translating the Thiemann operator into a spacetime covariant 
four-dimensional formalism, Reisenberger and Rovelli have noticed 
a suspicious lack of manifest 4-d covariance in the action of the 
operator \cite{ReisenbergerRovelli}, a fact pointing again to the 
possibility of anomalies in the quantum constraint algebra.

Motivated by these doubts, several variants of Thiemann's 
operator have been suggested.  The original Thiemann's operators 
is constructed using the volume operator.  There are two versions 
of the volume operator in the literature: $V_{RL}$, introduced in 
\cite{RovelliSmolin95} and $V_{AL}$, introduced in 
\cite{AshtekarLewandowski,AshtekarLewandowski2,%
AshtekarLewandowski3}.  See 
\cite{Lewandowski97} for a detailed comparison.  Originally, 
Thiemann thought that using $V_{RS}$ in the hamiltonian 
constraint would yield difficulties, but it later become clear 
that this is not the case \cite{LewandowskiMarolf}.  Both 
versions of the volume can be used in the definition, yielding 
two alternative versions of the hamiltonian 
\cite{LewandowskiMarolf}.  Next, in its simplest version the 
operator is non-symmetric.  Since the classical hamiltonian 
constraint constraint is real (on $SU(2)$ gauge invariant 
states), one might expect a corresponding self-adjoint quantum 
operator.\footnote{There is an argument which is often put 
forward against the requirement of self-adjointness of the 
hamiltonian constraint $H$: let $H$ be self-adjoint and $O$ be 
any operator of the form $O=[H,A]$ , where $A$ is any operator 
(many operators that we do not expect could vanish have this 
form).  Then the expectation value of $O$ vanishes on physical 
states $|\psi\rangle$ from 
$\langle\psi|O|\psi\rangle=\langle\psi|[H,A]|\psi\rangle=0$.  The 
mistake in this argument is easily detected by replacing $H$ with 
a nonrelativistic free particle quantum hamiltonian, $A$ with $x$ 
and $|\psi\rangle$ with an eigenstate of the momentum: the error 
is to neglect the infinities generated by the use of generalized 
states.} Accordingly, several ways of symmetrizing the operator 
have been considered (see a list in \cite{LewandowskiMarolf}).  
Next, Smolin has considered some ad hoc modifications of the 
constraint in \cite{Smolin96}.  Finally, the spacetime covariant 
formalism in \cite{ReisenbergerRovelli} naturally suggest a 
``covariantisation'' of the operator, described in 
\cite{ReisenbergerRovelli} under the name of ``crossing 
symmetry''.  This covariantisation amount to adding to the vertex 
described in Figure 3 the vertices, described in Figure 5, which 
are simply obtained by re-orienting Figure 3 in spacetime.
 \begin{figure}
 \centerline{\mbox{\epsfig{file=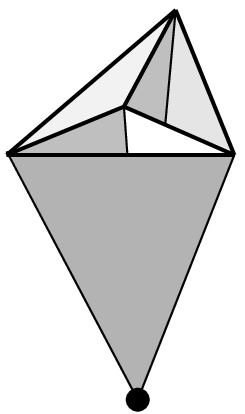}} \hskip1cm 
 \mbox{\epsfig{file=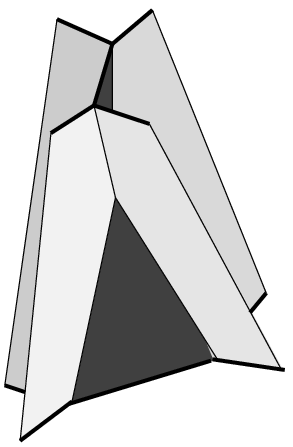}}}
 \caption{The crossing symmetric vertices.}
 \end{figure}

A full comparative analysis of this various proposals 
would be of great interest. 

Ultimately, the final tests of any proposal for the 
hamiltonian constraint operator must be consistency and a 
correct classical limit.  Thus, the solution of the 
hamiltonian constraint puzzle is likely to be subordinate 
to the solution of the problem of extracting the classical 
limit (of the dynamics) from the theory.
	
\item[Matter.] 

The basics of the description of matter in the loop 
formalism have been established in 
\cite{MoralesRovelli,MoralesRovelli2,KrasnovMatter,%
Baez,ThiemannFermions,ThiemannYM}.  Work needs to be done in order to 
develop a full description of the basic matter couplings.  In 
particular there are strong recurring indications that the Planck 
scale discreteness naturally cuts the traditional quantum field 
infinities off.  In particular, in \cite{ThiemannYM}, Thiemann argues 
that the Hamiltonian constraint governing the coordinate time 
evolution of the Yang-Mills field is a well defined operator (I recall 
that, due to the ultraviolet divergences, no rigorously well defined 
Hamiltonian operator for conventional Yang-Mills theory is known in 4 
dimensions.)  If these indications are confirmed, the result would be 
very remarkable.  What is still missing are calculational techniques 
that could allow us to connect the well-defined constraint with finite 
observables quantities such as scattering amplitudes.

\item[Spacetime formalism.]

 In my view, the development of continuous spacetime formalisms, 
 \cite{RovelliSurf,ReisenbergerRovelli,Baez97,SmolinMarkopoulo}, 
 is one of the most promising areas of development of the theory, 
 because it might be the key for addressing most of the open 
 problems.  First, a spacetime formalism frees us from the 
 obscurities of the frozen time formalism, and allows an 
 intuitive, Feynman-style, description of the dynamics of quantum 
 spacetime.  I think that the classical limit, the quantum 
 description of black holes, or graviton-graviton scattering, just 
 to mention a few examples, could be addressed much more easily in 
 the covariant picture.  Second, it allows the general ideas of 
 Hartle \cite{Hartle} and Isham \cite{Isham,Isham2,Isham3,Isham4} 
 on the interpretation of generally covariant quantum theories to 
 be applied in loop quantum gravity.  This could drastically 
 simplify the complications of the canonical way of dealing with 
 general covariant observables 
 \cite{RovelliObservables,RovelliObservables2}.  Third, the 
 spacetime formalism should suggest solutions to the problem of 
 selecting the correct hamiltonian constraint: it is usually 
 easier to deal with invariances in the lagrangian rather than in 
 the hamiltonian formalisms.  The spacetime formalism is just 
 born, and much has to be done.  See the original papers for 
 suggestions and open problems.

\item[Black holes.]

The derivation of the Bekenstein-Hawking entropy formula 
is a major success of loop quantum gravity, but much 
remains to be understood.   A clean derivation from the 
full quantum theory is not yet available.  Such a 
derivation would require us to understand what precisely 
is the event horizon in the quantum theory.  In other 
words, given a quantum state of the geometry, we should be 
able to define and ``locate'' its horizon (or whatever 
structure replaces it in the quantum theory).  To do so, 
we should understand how to effectively deal with the 
quantum dynamics, how to describe the classical limit (in 
order to find the quantum states corresponding to 
classical black hole solutions), as well as how do describe 
asymptotically flat quantum states.  

Besides these formal issues, at the roots of the black hole 
entropy puzzle there is a basic physical problem, which, to my 
understanding, is still open.  The problem is to understand how we 
can use basic thermodynamical and statistical ideas and techniques 
in a general covariant context.\footnote{A general approach to
this problem and an idea about its connection with the  
``problem of time'' in quantum gravity has been developed by Connes 
and Rovelli in \cite{RovelliStat1,RovelliStat2,ConnesRovelli}.} To 
appreciate the difficulty, notice that statistical mechanics makes 
heavy use of the notion of energy (say in the definition of the 
canonical or microcanonical ensembles); but there is no natural 
local notion of energy associated to a black hole (or there are 
too many of such notions).  Energy is an extremely slippery notion 
in gravity.  Thus, how do we define the statistical ensemble?  Put 
in other words: to compute the entropy (say in the microcanonical) 
of a normal system, we count the states with a given energy.  In 
GR we should count the states with a given {\em what\/}?  One may 
say: black hole states with a given area.  But why so?  We do 
understand why the number of states with given {\em energy\/} 
governs the thermodynamical behavior of normal systems.  But why 
should the number of states with given {\em area\/} govern the 
thermodynamical behavior of the system, namely govern its heat 
exchanges with the exterior?  A tentative physical discussion of 
this last point can be found in \cite{RovelliAscona}.

\item[How to extract physics from the theory.]
	
Assume we pick a specific hamiltonian constraint.  Then we 
have, in principle, a well defined quantum theory.  How do 
we extract physical information from it?  Some physical 
consequences of the theory, such as the area and volume 
eigenvalues, or the entropy formula, have been extracted 
from the theory by various ad hoc methods.  But is there a 
general technique, say corresponding to the traditional QFT 
perturbation expansion of the $S$ matrix, for describing 
the full dynamics of the gravitational field?  Presumably, 
such general technique should involve some kind of 
expansion, since we could not hope to solve 
the theory exactly.  Attempts to define physical 
expansions have been initiated in \cite{Rovelli95b} 
and, in different form, in \cite{ReisenbergerRovelli}.  
Ideally, one would want a general scheme for computing transition 
amplitudes in some expansion parameter around some 
state. Computing scattering amplitudes would be of 
particular interest, in order to make connection with 
particle physics language and to compare the theory with 
string predictions. 

\item[Classical limit.]

Finally, to prove that loop quantum gravity is a valuable 
candidate for describing quantum spacetime, we need to prove that 
its classical limit is GR (or at least overlaps GR in the regime 
where GR is well tested).  The traditional connection between loop 
quantum gravity and classical GR is via the notion of weave, a 
quantum state that ``looks semiclassical'' at distances large 
compared to the Planck scale.  However, the weaves studied so far 
\cite{weave,GrotRovelli} are 3d weaves, in the sense that they are 
eigenstates of the three dimensional metric.  Such a state 
corresponds to an eigenstate of the position for a particle.  
Classical behavior is recovered not by these states but rather by 
wave packets which have small spread in position as well as in 
momentum.  Similarly, the quantum Minkowski spacetime should have 
small spread in the three metric as well as in its momentum -- as 
the quantum electromagnetic vacuum has small quantum spread in the 
electric and magnetic field.  To recover classical GR from loop 
quantum gravity, we must understand such states.  Preliminary 
investigation in this direction can be found in 
\cite{IwasakiRovelli2,IwasakiRovelli}, but these papers are now 
several years old, and they were written before the more recent 
solidification of the basics of the theory.  Another direction 
consists in the direct study of coherent states in the state space 
of the theory.

\end{description}

As these brief notes indicate, the various open problems in loop 
quantum gravity are interconnected.  In a sense, loop quantum 
gravity grew aiming at the nonperturbative regime and the physical 
results obtained so far are in this regime.  The main issue is 
then to recover the long distance behavior of the theory.  That 
is, to study its classical limit and the dynamics of the low 
energy excitations over a semiclassical background.  Understanding 
this aspect of the theory would assure us that the theory we are 
dealing with is indeed a quantum theory of the gravitational 
field, would allow us to understand quantum black holes, would 
clarify the origin of infinities in the matter hamiltonians and so 
on.  Still in other words, what mostly needs to be understood is 
the structure of the (Minkowski) vacuum in loop quantum gravity.

\section{Short summary and conclusion}\label{8}

In this section, I very briefly summarize the state of loop quantum 
gravity and its main results.  The mathematics of the theory is 
solidly defined, and is understood from several alternative points of 
view.  Long standing problems such as the lack of a scalar product, 
the difficult of controlling the overcompletness of the loop basis and 
the problem of implementing the reality condition in the quantum 
theory have been successfully solved or sidestepped.  The kinematics is 
given by the Hilbert space $\cal H$, defined in Section 
\ref{hilbertspace}, which carries a representation of the basic 
operators: the loop operators (\ref{t00}-\ref{t11}).  A convenient 
orthonormal basis in $\cal H$ is provided by the spin network states, 
defined in Section \ref{states}.  The diffeomorphism invariant states 
are given by the s-knot states, and the structure and properties of 
the (diff-invariant) quantum states of the geometry are quite well 
understood (Section \ref{diffinvariance}).  These states give a 
description of quantum spacetime in terms of polymer-like excitations 
of the geometry.  More precisely, in terms of elementary excitations 
carrying discretized quanta of area.

The dynamics is coded into the hamiltonian constraint.  A well defined 
version of this constraint exists (see equation (\ref{lee})), and thus 
a consistent theory exists, but a proof that the classical limit of 
this theory is classical general relativity is still lacking.  Alternative 
versions of the hamiltonian constraint have been proposed and are 
under investigation.  In all these cases the hamiltonian has the 
crucial properties of acting on nodes only.  This implies that its 
action is naturally discrete and combinatorial.  This fact is possibly 
at the roots of the finiteness of the theory.  A large class of 
physical states which are exact solutions of the dynamics are given by 
s-knots without nodes; other exact states are related to knot theory 
invariants (Section \ref{technical}).

The theory can be extended to include matter, and there are strong 
indications that ultraviolet divergences do not appear.  A spacetime 
covariant version of the theory, in the form of a topological sum over 
surfaces is under development (Section \ref{spacetime}).

The main physical results derived so far from the theory is given 
by the explicit computation of the eigenvalues of area and volume, 
some of which are given in equations (\ref{spec}-\ref{v2}), and a 
derivation of the black hole entropy formula (Section \ref{spec}).  
The two main (related) open problems are to understand the 
description of the low energy regime within the theory and to 
choose the correct version of the hamiltonian constraint.

\subsection{Conclusion}\label{9} 

The history of quantum gravity is a sequel of moments of great 
excitement followed by bitter disappointments.  I distinctively 
remember, as a young student, listening to a very famous physicist 
announcing at a major conference that quantum gravity was solved (I 
think it was the turn of supergravity).  The list of theories that 
claimed to be final and have then ended up forgotten or superseded is 
a reason of embarrassment for the theoretical physics community, in my 
opinion.

In my view, loop quantum gravity is the best we can do so far in 
trying to understand quantum spacetime, from a nonperturbative, 
background-independent point of view.  Theoretically, we have reasons 
to suspect that this approach could represent a consistent quantum 
theory with the correct classical limit, but there are also some 
worrying contrary indications.  The theory yields a definite physical 
picture of quantum spacetime and definite quantitative predictions, 
but a systematic way of extracting physical information is still 
lacking.  Experimentally, there is no support to the theory, neither 
direct nor indirect.  The spectra of area and volume computed in the 
theory could or could not be physically correct.  I wish I could live 
long enough to know!

\section*{Acknowledgments}

I am especially indebted with Michael Reisenberger, Roberto DePietri 
Don Marolf, Jerzy Lewandowski, John Baez, Thomas Thiemann and Abhay 
Ashtekar for their accurate reading of the manuscript, their detailed 
suggestions and their ferocious criticisms.  Their inputs have been 
extensive and their help precious.  I am grateful to all my friends in 
the community for the joy of doing physics together.  This work was 
supported by NSF Grant PHY-95-15506.

\bibliography{RovelliLivrev}

\end{document}